\documentclass[12pt]{article}
\usepackage{epsfig}
\makeatletter
\@addtoreset{equation}{section}

\renewcommand{\thefootnote}{\fnsymbol{footnote}}
\makeatother
\begin{document}
\begin{flushright}
{\tt hep-th/0602277} \\
HIP-2006-10/TH\\ 
TIT/HEP-551 \\
Feb., 2006
\end{flushright}
\vspace{2mm}
\begin{center}
{\large \bf Hyper-K{\"a}hler Sigma Models on 
(Co)tangent Bundles
\\ with $SO(n)$ Isometry}
\end{center}
\begin{center}
\normalsize
{\large \bf  Masato Arai$^a$
\footnote{E-mail: masato.arai@helsinki.fi}
and Muneto Nitta$^b$
\footnote{E-mail: nitta@th.phys.titech.ac.jp}  
}
\end{center}
\vskip 1.2em
\begin{center}
{\it 
$^a$High Energy Physics Division, 
             Department of Physical Sciences,
             University of Helsinki \\
 and Helsinki Institute of Physics,
 P.O.Box 64, FIN-00014, Finland \\
\vskip 1.0em
$^b$Department of Physics, Tokyo Institute of 
  Technology \\
  Tokyo 152-8551, Japan
\vskip 1.0em
}
\end{center}
\vskip 1.0cm
\begin{center}
{\large Abstract}
\vskip 0.7cm
\begin{minipage}[t]{14cm}
\baselineskip=19pt
\hskip4mm
We construct ${\cal N}=2$ supersymmetric 
nonlinear sigma models 
whose target spaces are 
tangent as well as cotangent bundles
 over the quadric surface 
 $Q^{n-2} = SO(n)/[SO(n-2)\times U(1)]$. 
 We use the projective superspace framework,
 which is an off-shell formalism of ${\cal N}=2$ supersymmetry.

%%%%%%%%%%%%%%%
\end{minipage}
\end{center}

%%%%%%  user's commands  %%%%%%%%%%%%%%%%%%%%%%%%%%%%%%%%%%%%%%%%%%%
\newcommand {\non}{\nonumber\\}
\newcommand {\eq}[1]{\label {eq.#1}}
\newcommand {\defeq}{\stackrel{\rm def}{=}}
\newcommand {\gto}{\stackrel{g}{\to}}
\newcommand {\hto}{\stackrel{h}{\to}}
\newcommand {\1}[1]{\frac{1}{#1}}
\newcommand {\2}[1]{\frac{i}{#1}}
\newcommand{\be}{\begin{eqnarray}}
\newcommand{\ee}{\end{eqnarray}}
\newcommand {\thb}{\bar{\theta}}
\newcommand {\ps}{\psi}
\newcommand {\psb}{\bar{\psi}}
\newcommand {\ph}{\varphi}
\newcommand {\phs}[1]{\varphi^{*#1}}
\newcommand {\sig}{\sigma}
\newcommand {\sigb}{\bar{\sigma}}
\newcommand {\Ph}{\Phi}
\newcommand {\Phd}{\Phi^{\dagger}}
\newcommand {\Sig}{\Sigma}
\newcommand {\Phm}{{\mit\Phi}}
\newcommand {\eps}{\varepsilon}
\newcommand {\del}{\partial}
\newcommand {\dagg}{^{\dagger}}
\newcommand {\pri}{^{\prime}}
\newcommand {\prip}{^{\prime\prime}}
\newcommand {\pripp}{^{\prime\prime\prime}}
\newcommand {\prippp}{^{\prime\prime\prime\prime}}
\newcommand {\pripppp}{^{\prime\prime\prime\prime\prime}}
\newcommand {\delb}{\bar{\partial}}
\newcommand {\zb}{\bar{z}}
\newcommand {\mub}{\bar{\mu}}
\newcommand {\nub}{\bar{\nu}}
\newcommand {\lam}{\lambda}
\newcommand {\lamb}{\bar{\lambda}}
\newcommand {\kap}{\kappa}
\newcommand {\kapb}{\bar{\kappa}}
\newcommand {\xib}{\bar{\xi}}
\newcommand {\ep}{\epsilon}
\newcommand {\epb}{\bar{\epsilon}}
\newcommand {\Ga}{\Gamma}
\newcommand {\rhob}{\bar{\rho}}
\newcommand {\etab}{\bar{\eta}}
\newcommand {\chib}{\bar{\chi}}
\newcommand {\tht}{\tilde{\th}}
\newcommand {\zbasis}[1]{\del/\del z^{#1}}
\newcommand {\zbbasis}[1]{\del/\del \bar{z}^{#1}}
\newcommand {\vecv}{\vec{v}^{\, \prime}}
\newcommand {\vecvd}{\vec{v}^{\, \prime \dagger}}
\newcommand {\vecvs}{\vec{v}^{\, \prime *}}
\newcommand {\alpht}{\tilde{\alpha}}
\newcommand {\xipd}{\xi^{\prime\dagger}}
\newcommand {\pris}{^{\prime *}}
\newcommand {\prid}{^{\prime \dagger}}
\newcommand {\Jto}{\stackrel{J}{\to}}
\newcommand {\vprid}{v^{\prime 2}}
\newcommand {\vpriq}{v^{\prime 4}}
\newcommand {\vt}{\tilde{v}}
\newcommand {\vecvt}{\vec{\tilde{v}}}
\newcommand {\vecpht}{\vec{\tilde{\phi}}}
\newcommand {\pht}{\tilde{\phi}}
\newcommand {\goto}{\stackrel{g_0}{\to}}
\newcommand {\tr}{{\rm tr}\,}
\newcommand {\GC}{G^{\bf C}}
\newcommand {\HC}{H^{\bf C}}
\newcommand{\vs}[1]{\vspace{#1 mm}}
\newcommand{\hs}[1]{\hspace{#1 mm}}
\newcommand{\al}{\alpha}
\newcommand{\Lam}{\Lambda}
\newcommand{\changed}[1]{{\bf #1}}

\newpage

\setcounter{footnote}{0}
\renewcommand{\thefootnote}{\arabic{footnote}}
\section{Introduction}
Hyper-K\"ahler manifolds provide 
 a fruitful relation
 between physics and mathematics. 
One of ingredients to study them may be solitons. 
For instance it is well-known that 
 the moduli space for Yang-Mills (YM) instantons 
 on ${\bf R}^4$~\cite{ADHM} and one of 
 Bogomol'nyi-Prasad-Sommerfield 
(BPS) monopoles~\cite{Nahm,Atiyah:1985dv}
 are both hyper-K\"ahler. 
Gravitational instantons \cite{grav-instanton,Kronheimer:1989zs} 
 and the moduli space of YM instantons on 
 gravitational instantons~\cite{KN} are also hyper-K\"ahler. 
More direct connection related to these 
would be given by supersymmetric nonlinear sigma models 
with eight supercharges 
(like ${\cal N}=2$ supersymmetry in $d=4$ 
and ${\cal N}=(4,4)$ supersymmetry in $d=2$): 
scalar fields in these models 
belong to hypermultiplets, 
parametrizing target spaces which 
must be hyper-K\"ahler from the requirement of 
${\cal N}=2$ supersymmetry \cite{Alvarez-Gaume:1981hm}. 
Conversely there exists the unique (massless) 
nonlinear sigma model 
for arbitrary hyper-K\"ahler manifold. 
Hyper-K\"ahler structure on soliton moduli spaces 
can be understood in terms of nonlinear sigma models as follows. 
Instantons and BPS monopoles can be naturally embedded into 
supersymmetric gauge theories with sixteen supercharges, 
and they preserve/break half of supersymmetry. 
Their dynamics can be described 
by nonlinear sigma models with 
eight supersymmetry preserved by them.  
The hyper-K\"ahler quotient 
was discovered in the context of 
nonlinear sigma models \cite{Lindstrom:1983rt,Hitchin:1986ea}. 
Since then it has become an important tool: 
YM instantons, gravitational instantons and 
YM instantons on gravitational instantons 
can be obtained by certain hyper-K\"ahler quotients 
\cite{ADHM,Kronheimer:1989zs,KN}.  
Hyper-K\"ahler sigma models also
give low energy effective action on 
the Higgs branch of vacua in ${\cal N}=2$ supersymmetric gauge theories 
\cite{Argyres:1996eh,Antoniadis:1996ra}, 
where target space is obtained as
the hyper-K\"ahler quotient.

Construction of explicit metrics on 
hyper-K\"ahler manifolds 
is an important problem. 
Compact hyper-K\"ahler manifolds are difficult to construct 
whereas known ones are all non-compact. 
In general an isometry of manifolds often restricts 
the form of their metrics. 
An important class of hyper-K\"ahler manifolds 
is given by toric hyper-K\"ahler (hypertoric)
manifolds~\cite{Lindstrom:1983rt,Pedersen:1988vw,Gibbons:1996nt,Bielawski,Eto:2005wf}, 
namely $4n$ dimensional hyper-K\"ahler manifolds
admitting mutually commuting $n$ tri-holomorphic isometries. 
This class of manifolds was firstly found in 
construction of the general action of ${\cal N}=2$ 
supersymmetric tensor multiplets \cite{Lindstrom:1983rt}, 
which can be dualized by the Legendre transformation to 
hypermultiplets with toric hyper-K\"ahler manifolds.
Thus ${\cal N}=2$ supersymmetric models are obviously useful to 
study hyper-K\"ahler manifolds.
Any toric hyper-K\"ahler manifold (with dimension $4n$) can be 
obtained as hyper-K\"ahler quotient of flat space ${\bf H}^{n+m}$
by $U(1)^{m}$ \cite{Bielawski}. 
Hyper-K\"ahler manifolds which are not toric were studied 
 with hyper-K\"ahler quotient by a product of gauge group
 $\Pi_{i=1}^kU(N_i)$ in Ref. \cite{Lindstrom:1999pz}.

An isometry
of non-Abelian group $G$ is further restrictive 
for metrics of manifolds.
Homogeneous K\"ahler manifolds $G/H$ 
were completely classified \cite{Borel,Bordemann:1985xy}
and their K\"ahler potentials 
were systematically constructed \cite{Itoh:1985ha}. 
Hyper-K\"ahler manifolds cannot become homogeneous, 
so we may consider in a slightly different way.
Let us remember that 
homogeneous K\"ahler manifolds can be formulated as 
co-adjoint orbits of Lie algebra $G$ 
with the so-called Kirillov-Kostant-Souriau symplectic structure \cite{Be}. 
Then, co-adjoint orbits of complex Lie algebra $G^{\bf C}$ 
become cotangent bundles  
$G^{\bf C}/ H^{\bf C} \simeq T^* (G/H)$  
over homogeneous K\"ahler manifolds $G/H$, 
and they are known to 
admit hyper-K\"ahler metrics \cite{Kr2}.\footnote{Stenzel constructed 
the Ricci-flat K\"ahler metric on 
complexification of Riemann symmetric spaces 
$G^{\bf C}/ H^{\bf C} \simeq T^* (G/H)$~\cite{Stenzel}. 
See \cite{Cvetic:2000db,Higashijima:2001yn} for explicit metric 
in the case of $S^{N-1} \simeq SO(N)/SO(N-1)$.
} 
More explicit analysis was performed
for 
cotangent bundle over Hermitian symmetric spaces~\cite{BG}.
Later it was shown that 
cotangent bundles over any K\"ahler manifolds $M$ (without any isometry) admit 
hyper-K\"ahler metrics at least in neighbour of $M$~\cite{Kaledin,Feix}. 

When one would like to construct 
arbitrary hyper-K\"ahler manifold, 
fully off-shell ${\cal N}=2$ supersymmetric 
formalisms should be useful. 
The harmonic superspace provides such 
a fully off-shell ${\cal N}=2$ superspace \cite{harmonic1,harmonic2}. 
It can provide the most general action for hypermultiplets
which induces the most general hyper-K\"ahler manifolds 
if one can eliminate infinite number of auxiliary fields. 
The projective superspace \cite{karlhede}--\cite{gon} 
is another fully off-shell ${\cal N}=2$ superspace. 
Its equivalence to harmonic superspace was discussed \cite{Kuzenko:1998xm}. 
Recently the projective superspace in five- and six-dimensions has also 
 been studied \cite{Kuzenko:2005sz,Gates:2005mc,Kuzenko:2006mv}.  
In the six-dimensional case, the projective superspace was first 
 introduced in \cite{Grundberg:1984xr} 
 to construct self-coupling of ${\cal N}=(1,0)$ tensor multiplets.

It was shown by Gates and Kuzenko \cite{kuzenko,kuzenko2} that 
the particular multiplets 
in the projective superspace, called 
the {\it polar multiplets},   
are suitable to describe 
${\cal N}=2$ supersymmetric nonlinear sigma models on 
{\it tangent} (but not cotangent) bundles 
$TM$ over K\"ahler manifolds $M$.\footnote{
They also studied hyper-K{\"a}hler manifolds by using c-map in the
 projective superspace framework \cite{Gates:1999zv}.}
The polar multiplets $\Upsilon^i$ contain chiral superfields 
 $\Phi^i$ and complex linear (nonminimal) superfields 
 $\Sigma^i$ \cite{Gates:1983nr} 
 in terms of ${\cal N}=1$ superfields, where  
$\Phi^i$ parametrize the base K\"ahler manifold $M$  
and $\Sigma^i$ are regarded as components of 
a tangent vector on $M$. 
Complex linear superfields $\Sigma^i$ 
can be dualized by the Legendre transformation 
to chiral superfields $\psi_i$ which can be 
regarded as components of a cotangent vector. 
Then, the nonlinear sigma models on cotangent bundles $T^*M$ 
in terms of solely chiral superfields $(\Phi^i,\psi_i)$ are obtained. 
Once the K\"ahler potential $K(\Phi,\Phi^\dagger)$ 
of ${\cal N}=1$ supersymmetric model 
on any K\"ahler manifold $M$ is given, 
one can easily obtain its ${\cal N}=2$ supersymmetric extension 
on (co)tangent bundle $T^{(*)}M$
with replacing chiral superfields $\Phi^i$ 
by polar multiplets $\Upsilon^i$. 
This is nicely conforming to the mathematical result
\cite{Kaledin,Feix}.\footnote{The work of
 Gates and Kuzenko was done independently from \cite{Kaledin,Feix}.
 In fact it was earlier than \cite{Feix}. }
However, the main problem to obtain explicit action in terms of 
component fields (or ${\cal N}=1$ superfields) 
is that one has to eliminate 
infinite number of auxiliary ${\cal N}=1$ superfields 
contained in the polar multiplets. 
The authors in \cite{kuzenko,kuzenko2} 
explicitly constructed 
nonlinear sigma models on tangent and cotangent bundles 
over the complex projective space 
${\bf C}P^{n-1} = SU(n)/[SU(n-1) \times U(1)]$, 
 which is one of the Hermitian symmetric spaces, 
 by eliminating auxiliary fields 
 with the help of the isometry $SU(n)$ on ${\bf C}P^{n-1}$. 
The cotangent bundle action recovers the $T^*{\bf C}P^{n-1}$ sigma model
 constructed by the hyper-Kahler quotients 
 \cite{Curtright:1979yz}.
The purpose of the present paper is to construct 
 ${\cal N}=2$ supersymmetric nonlinear sigma models on 
 (co)tangent bundle 
 on another Hermitian symmetric space, 
 the so-called {\it quadric surface} $Q^{n-2} = SO(n)/[SO(n-2)\times U(1)]$, 
 with following their work.

This paper is organized as follows. 
We give a review of the K\"ahler quotient construction of 
$Q^n$ \cite{Higashijima:1999ki,Higashijima:2000rp} 
in the rest of introduction. 
In section 2, we give a brief review of the projective superspace. 
In section 3, we review how to construct nonlinear sigma models with
 tangent and cotangent bundles over the projective space ${\bf C}P^{n-1}$.
We consider a sigma model with tangent bundle $TQ^{n-2}$ in section 4.
In subsection 4.1, by using the isomorphism 
 $Q^2\simeq {\bf C}P^1\times {\bf C}P^1$,
 we construct the sigma model
 with tangent bundle $TQ^{2}$.
In subsection 4.2, we solve the equations of motion for auxiliary fields 
 and derive the $TQ^{n-2}$ action.
In section 5, we derive the nonlinear sigma 
 action with cotangent bundle $T^*Q^{2}$ via the Legendre
 transformation.
Further we propose the the cotangent bundle action 
 for $T^*Q^{n-2}$.
Section 6 is devoted to discussion. 
In Appendix A, we review an another method to eliminate infinite set of 
 auxiliary fields based on
 the duality between polar and $O(2)$ multiplet with some examples.
In Appendix B,
 we show the detailed derivation of 
 identities between metric and Riemann tensor.
In Appendix C,
 we derive the sigma model with cotangent bundle $T^*Q^2$
 with the isomorphism $T^*Q^2\simeq T^*{\bf C}P^1\times T^*{\bf C}P^1$.
We discuss the $n=3$ case in $T^{(*)}Q^{n-2}$ sigma model, and show that
 the solution for $\Upsilon$ and the (co)tangent bundle action
 is $T^{(*)}{\bf C}P^1$'s one in Appendix D.

Before closing introduction we 
review how to construct 
nonlinear sigma models on the quadric surface $Q^{n-2}$ 
in terms of the ${\cal N}=1$ superfields 
\cite{Higashijima:1999ki,Higashijima:2000rp}. 
Let $\phi^i(x,\theta,\bar \theta)$ 
($i=1,\cdots,n$) 
be chiral superfields, 
$\bar D_{\dot \alpha} \phi^i = 0$
belonging to the vector representation of $SO(n)$. 
Introducing an auxiliary vector superfield 
$V (x,\theta,\bar \theta) \; (= V^\dagger)$
and an auxiliary chiral superfield  $\sigma (x,\theta,\bar \theta)$ 
($\bar D_{\dot \alpha} \sigma = 0$), 
 being a singlet representation of $SO(n)$,
 the Lagrangian can be written as  
\begin{eqnarray}
 {\cal L} = \int d^4 \theta 
 (\phi^{i\dagger}\phi^i e^V - r^2 V)
 + \left( \int d^2 \theta \, \sigma \phi^i \phi^i + {\rm c.c.}\right)
 \label{Q^n-lag}
\end{eqnarray}
with summation over repeated index $i$ implied, 
and $r^2$ a real positive constant called Fayet-Iliopoulos parameter. 
This Lagrangian possesses gauge invariance
\begin{eqnarray}
 V \to V - \Lambda - \Lambda^\dagger , \quad
 \phi^i \to e^{\Lambda} \phi^i, \quad
 \sigma \to e^{-2 \Lambda} \sigma 
\end{eqnarray}
with arbitrary chiral superfield $\Lambda(x,\theta,\bar \theta)$. 
Equation of motion of $V$ read $\phi^{i\dagger}\phi^i e^V -r^2 =0$, 
which can be solved as $V = - \log (\phi^{i\dagger}\phi^i/r^2)$. 
When the superpotential is absent in the Lagrangian (\ref{Q^n-lag}), 
we obtain the K\"ahler potential of ${\bf C}P^{n-1}$ 
by substituting the solution back into the K\"ahler potential of 
(\ref{Q^n-lag}) as  
\begin{eqnarray}
  K = r^2 \log \left(1 + {|\Phi^i|^2 \over r^2}\right) 
\end{eqnarray}
with a gauge fixing $\vec{\phi} = (\Phi^i,r)$ ($i=1,\cdots,n-1$). 
But now there exists the superpotential in the Lagrangian
 (\ref{Q^n-lag}).
Decomposing  $\phi^i$ in the representation of 
the $SO(n-2)\times U(1)$ group of $SO(n)$ 
 as $\phi^i=(x,y^j,z)~(j=1,\cdots,n-2)$,
 the $SO(n)$ transformation law is given by \cite{Higashijima:1999ki}
\begin{eqnarray}
 \delta\phi^i=
 \left(
 \begin{array}{ccc}
  i\theta & \bar{\epsilon}^{\bar{j}} & 0 \\
  -\epsilon^i & \theta^{ij} & -\bar{\epsilon}^{\bar{i}} \\
  0 & \epsilon^j & -i\theta
 \end{array}
 \right)
 \left(
 \begin{array}{c}
  x \\
  y^j \\
  z
 \end{array}
 \right)\,, \label{trans_SO}
\end{eqnarray}
where $\theta^{ij}={i \over 2}\theta^{kl}(T^{kl})^{ij}$ and
 $(T^{ij})^{kl}
 ={1 \over i}(\delta^{ik}\delta^{jl}-\delta^{kj}\delta^{il})$.
We take the rank-2 invariant tensor as
\begin{eqnarray}
J = \left(
    \begin{array}{ccc}
        0 &  {\bf 0}      &1 \\
 {\bf  0} & {\bf 1}_{n-2} &{\bf 0} \\
        1 &  {\bf 0}      &0      
    \end{array}
            \right) .
\end{eqnarray}
The equation of motion of $\sigma$ gives the constraint 
\begin{eqnarray}
 \vec{\phi}^2 = \phi^T J \phi = 2xz+y^2=0\,.
\end{eqnarray}
This can be solved to give 
\begin{eqnarray}
 \vec{\phi}=\left(
             \begin{array}{c}
	      x \\
              y^j \\
              -{y^2 \over 2x}
	     \end{array}
            \right). \label{so-vec}
\end{eqnarray}
With a gauge fixing 
$x=r$,
$\vec{\phi}^T = (r,\Phi^i, - {1\over 2r}\Phi^2)$, 
%$x=1$, 
%$\vec{\phi}^T = (1,\Phi^i, - {1\over 2} \Phi^2)$, 
we obtain the K\"ahler potential of the quadric surface 
\cite{delduc,Higashijima:1999ki,Higashijima:2000rp,Higashijima:2001de}, 
given by 
\begin{eqnarray}
  K(\Phi^i,\bar{\Phi}^{\bar{j}})
 =r^2\ln\left(1+{|\Phi^i|^2 \over r^2}
  +{(\Phi^i)^2(\bar{\Phi}^{\bar{j}})^2 \over 4r^4}\right)\,.
 \label{kahler_SO}
\end{eqnarray}
The K{\"a}hler metric can be calculated %, 
to give 
\begin{eqnarray}
 g_{i{\bar{j}}}&=&
 {\partial^2 K \over \partial \Phi^i \partial \bar{\Phi}^{\bar{j}}}
 =
  {\delta_{i\bar{j}} \over 1+{|\Phi^k|^2 \over r^2}
    +{(\Phi^l)^2(\bar{\Phi}^{\bar{m}})^2 \over 4r^4}}
 +{{\Phi^i\bar{\Phi}^{\bar{j}}-\Phi^j\bar{\Phi}^{\bar{i}} \over r^2}
   +{2\Phi^{i}\bar{\Phi}^{\bar{j}}|\Phi^k|^2 - \Phi^i\Phi^j(\bar{\Phi}^k)^2
    - \bar{\Phi}^i\bar{\Phi}^{\bar{j}}(\Phi^k)^2 \over 2r^4}
  \over \left(1+{|\Phi^l|^2 \over r^2}
  +{(\Phi^m)^2(\bar{\Phi}^{\bar{n}})^2 \over 4r^4}\right)^2}\,.\label{metric_quad}
\end{eqnarray}

%%%%%%%%%%%%%%%%%%%%%%%%%%%%%%%%%%%%%%%%%%%%%%%%%%%%%%%%%
\section{Projective Superspace}
The projective superspace \cite{karlhede}--\cite{gon} 
 consists of a complex projective 
 coordinate $\zeta$, which is an
 inhomogeneous coordinate of ${\bf C}P^1$, 
 in addition to ${\cal N}=2$ global superspace ${\bf R}^{4|8}$ 
 parameterized by 
\begin{eqnarray}
 z^M=(x^\mu,\theta^{i\alpha},\bar{\theta}^i_{\dot{\alpha}})\,,~~~~~
 \overline{\theta_i^\alpha}=\bar{\theta}^{\dot{\alpha}i}\,,~~~~
 i=1,2\,
 \label{full}
\end{eqnarray}
where the index $i$ labels the fundamental representation of 
the automorphism group $SU(2)_R$. 
Superfields $\Upsilon$ on the projective superspace 
are functions of this projective superspace %, and they are, however, 
with the constraints 
\begin{eqnarray}
 \nabla_\alpha \Upsilon(z,\zeta)
 =\bar{\nabla}_{\dot{\alpha}}\Upsilon(z,\zeta)=0\,,\label{const1}
\end{eqnarray}
where $\nabla_{\alpha}$ and $\bar{\nabla}_{\dot{\alpha}}$ are linear
 combination of ${\cal N}=2$ supercovariant derivatives in four dimensions, 
 given by  
\begin{eqnarray}
 \nabla_{\alpha}(\zeta)=D_{1\alpha}+\zeta D_{2\alpha}\,,~~~~
 \bar{\nabla}_{\dot{\alpha}}(\zeta)=\bar{D}^2_{\dot{\alpha}}
 -\zeta \bar{D}^1_{\dot{\alpha}}\,. \label{cov1}
\end{eqnarray}
Here the supercovariant derivatives satisfy the following 
 algebra \footnote{We take the normalization as 
 $D^2={1 \over 4}D^\alpha D_{\alpha}$.}
\begin{eqnarray}
\{D_{i\alpha},D_{j\beta}\}
 =\{\bar{D}_{i\dot{\alpha}},\bar{D}_{j\dot{\beta}}\}=0\,,~~~
\{D_{i\alpha},\bar{D}^j_{\dot{\beta}}\}
 =-2i\delta^j_i\partial_{\alpha\dot{\beta}}\,.
\end{eqnarray}
Notice that $\bar{\nabla}_{\dot{\alpha}}$ is the conjugate of 
 $\nabla_\alpha$
 under 
 the composition of complex conjugation with the antipodal map on
 the Riemann sphere, $\bar{\zeta}\rightarrow -1/\zeta$, 
 and multiplication by an appropriate factor.
For example,
\begin{eqnarray}
 \bar{\nabla}_{\dot{\alpha}}(\zeta)
 =(-\zeta)(\nabla_\alpha)^*\left(-{1 \over \zeta}\right)\,.
\end{eqnarray}
In the following, all conjugate of fields and operators in projective
 superspace are defined in this sense.

The constraints (\ref{const1}) for superfields are analogous to one
 for a chiral superfield in ${\cal N}=1$ superspace formalism 
 where the chiral subspace is defined. 
 The constraints (\ref{const1})
 define a subspace of the full ${\cal N}=2$ superspace (\ref{full}).
Since a function $K(\Upsilon,\bar{\Upsilon})$ 
of superfields is independent of 
some (a half) of the Grassmann coordinates of ${\cal N}=2$
superspace by definition (\ref{const1}), its integration over the  
 orthogonal operators for (\ref{cov1}).
\begin{eqnarray}
 \Delta_\alpha=\zeta^{-1}D_{1\alpha}-D_{2\alpha}\,,~~~
 \bar{\Delta}_{\dot{\alpha}}=\zeta^{-1}\bar{D}_{\dot{\alpha}}^2
 +\bar{D}_{\dot{\alpha}}^1\,\label{orth}
\end{eqnarray}
is invariant under ${\cal N}=2$ supersymmetry.
This leads to the following ${\cal N}=2$ invariant action:
\begin{eqnarray}
 S={1 \over 32\pi i}\int d^4x \oint_C \zeta d\zeta
  {\Delta^2\bar{\Delta}^2 \over 16}
 K(\Upsilon,\bar{\Upsilon},\zeta)\,.\label{action1}
\end{eqnarray}
Here the integration contour $C$ in the $\zeta$-plane 
 is supposed to be chosen to make the action (\ref{action1}) 
 nontrivial (i.e. not equal to zero).
In the following, 
we take the contour to surround 
the origin in the $\zeta$-plane.  

The algebra for $\nabla,\bar{\nabla},\Delta$ and $\bar{\Delta}$ is
given by 
\begin{eqnarray}
 &\{\nabla,\nabla\}=\{\nabla,\bar{\nabla}\}=\{\Delta,\Delta\}
 =\{\Delta,\bar{\Delta}\}=\{\nabla,\Delta\}=0\,,& \label{alg1} \\
 &\{\nabla_{\alpha},\bar{\Delta}_{\dot{\alpha}}\}
 =-\{\bar{\nabla}_{\dot{\alpha}},\Delta_{\alpha}\}
 =4i\partial_{\alpha\dot{\alpha}}\,.& \label{alg2}
\end{eqnarray}
Using Eqs.~(\ref{const1}), (\ref{alg1}) and (\ref{alg2}) 
with the identities
\begin{eqnarray}
 \Delta_\alpha=\zeta^{-1}(2D_{\alpha}-\nabla_\alpha),~~~~
 \bar{\Delta}_{\dot{\alpha}}=2\bar{D}_{\dot{\alpha}}
 +\zeta^{-1}\bar{\nabla}_{\dot{\alpha}}\,,
\end{eqnarray} 
the manifestly ${\cal N}=2$ supersymmetric 
action (\ref{action1}) can be reduced to 
the action in terms of ${\cal N}=1$ superfields,  
\begin{eqnarray}
 S=\int d^4x {1 \over 2\pi i}\oint_C {d\zeta \over \zeta}
 {D^2 \bar{D}^2 \over 16}K(\Upsilon |,\bar{\Upsilon}|,\zeta)
  =\int d^8z {1 \over 2\pi i}\oint_C {d\zeta \over \zeta}
  K(\Upsilon |,\bar{\Upsilon}|,\zeta)\,  \label{action2}
\end{eqnarray}
with ${\cal N}=1$ superfield covariant derivative defined by 
 $D_{\alpha} \equiv D_{1\alpha}$, and $d^8z \equiv
 d^4xD^2\bar{D}^2/16=d^4xd^2\theta d^2\bar{\theta}$.
Here $\Upsilon{|}$ indicates the $\theta^2$ and $\bar{\theta}^2$
 independent part of a superfield $\Upsilon$.
In the following, we will simply write it as $\Upsilon$. 

The superfields obeying the constraints (\ref{const1})
 are classified into 
 (i) real/complex $O(k)$ multiplets \cite{Ketov:1987yw}, 
 (ii) rational multiplets \cite{rocek}, 
 and (iii) analytic multiplets \cite{rocek}.
Furthermore analytic multiplets 
contain the so-called polar multiplets and 
the real tropical multiplets, 
which describe charged
 ${\cal N}=2$ hypermultiplets and vector multiplets, 
 respectively \cite{gon0,gon}.\footnote{
 Cutting off the power series in (\ref{polar}) at some finite
 $k (>2)$, one results in the complex $O(k)$ multiplet. The case
 $k=1$ corresponds to the on-shell hypermultiplet, while for $k=2$ we
 obtain two tensor multiplets.}
In what follows,
 we focus on the polar multiplets to consider 
 ${\cal N}=2$ supersymmetric nonlinear sigma models.
The polar multiplets and their conjugation 
can be expanded in terms of $\zeta$ as 
\begin{eqnarray}
 \Upsilon(z,\zeta)=\sum_{n=0}^\infty\Upsilon_n(z)\zeta^n,
~~~~~
 \bar \Upsilon(z,\zeta)=\sum_{n=0}^\infty\bar{\Upsilon}_n(z)
 \left(-{1 \over \zeta} \right)^n \,, \label{polar}
\end{eqnarray}
respectively.\footnote{
The projective superfields $\Upsilon$ and $\bar{\Upsilon}$ are called
 arctic and antarctic \cite{gon0,gon}, respectively.
}
Here all $\Upsilon_n$ (and $\bar \Upsilon_n$) 
are ${\cal N}=1$ superfields:  
 $\Upsilon_0$ is a chiral superfield, 
 $\Upsilon_1$ 
 a complex linear (or nonminimal) superfield \cite{Gates:1983nr}, 
satisfying the ${\cal N}=1$ constraints 
\begin{eqnarray}
 \bar{D}_{\dot{\alpha}}\Upsilon_0= 0, \quad 
 \bar{D}^2\Upsilon_1=0\,,
\end{eqnarray}
respectively, due to the constraints (\ref{const1}). 
The rests of fields $\Upsilon_2$, $\Upsilon_3,\dots,$ 
are complex unconstrained superfields, 
which are always auxiliary once the action is given.

The free action obeying hermiticity and ${\cal N}=2$ supersymmetry
 is given by
\begin{eqnarray}
 S_{free}=\int d^8z\oint_C {d\zeta \over 2\pi i \zeta}
 \bar{\Upsilon}\Upsilon\,.
\end{eqnarray}
On the other hand, 
the action (\ref{action2}) with the polar multiplets $\Upsilon$ 
is the most general action for  
${\cal N}=2$ supersymmetric nonlinear sigma models 
on the tangent (but not cotangent) 
bundles over K\"ahler manifolds. 
For convenience 
let us rewrite physical ${\cal N}=1$ superfields in 
the projective superfields $\Upsilon^i$ 
(with $i$ labelling projective superfields) as
\begin{eqnarray}
 \Phi^i \equiv \Upsilon^i (\zeta){\Bigg |}_{\zeta=0} \,,~~~~~~
 \Sigma^i \equiv {d \Upsilon^i(\zeta) \over d\zeta}{\Bigg |}_{\zeta=0} .
\end{eqnarray}
Then, $\Phi^i$ and $\Sigma^i$ are regarded as 
coordinates of the base K\"ahler manifold 
and components of a tangent vector, respectively,  
as explained as follows. 
The action (\ref{action2}) respects all the geometric features 
which ${\cal N}=1$ supersymmetric 
 nonlinear sigma model on K\"ahler manifolds possesses.
For instance, 
the action of ${\cal N}=1$ supersymmetric nonlinear sigma model 
\begin{eqnarray}
 S=\int d^8z K(\Phi^i,\bar{\Phi}^{\bar i})\, 
 \label{kahler1}
\end{eqnarray}
is invariant under the K{\"a}hler transformation
\begin{eqnarray}
 K(\Phi,\bar{\Phi}) 
 \rightarrow 
 K(\Phi,\bar{\Phi})+(\Lambda(\Phi)
     +\bar{\Lambda}(\bar{\Phi}))\,. \label{kahler-trans}
\end{eqnarray}
This invariance can be promoted to 
\begin{eqnarray}
 K(\Upsilon,\bar{\Upsilon})\rightarrow K(\Upsilon,\bar{\Upsilon})
 + (\Lambda(\Upsilon)+\bar{\Lambda}(\bar{\Upsilon}))\,,
\end{eqnarray}
for the action (\ref{action2}).
A holomorphic field redefinition 
$\Phi^i \rightarrow f^i(\Phi^j)$ of the chiral superfields 
in the action (\ref{kahler1}) gives 
a holomorphic coordinate transformation. 
% where the indices $i,j$ denote flavor.
This is promoted for the action (\ref{action2}) to 
\begin{eqnarray}
  \Upsilon^i \rightarrow f^i(\Upsilon^j)\, 
\end{eqnarray}
deducing the transformation laws of $\Phi^i$ and $\Sigma^i$
as holomorphic coordinates 
of the base K{\"a}hler manifold, 
$\Phi^i \rightarrow f^i(\Phi^j)$, 
and 
components of a holomorphic tangent vector, 
$\Sigma^i \rightarrow {{\partial f^i \over \partial \Phi^j}} (\Phi) \Sigma^j$.  
%at the point $\Phi$ of the same manifold.
Thus, the set of fields $(\Phi^i,\Sigma^i)$ parameterizes 
the tangent bundles of the K{\"a}hler manifolds.
Note that the action (\ref{action2})
 is invariant under the rigid $U(1)$ transformations
\begin{eqnarray}
 \Upsilon(\zeta)\rightarrow \Upsilon(e^{i\alpha}\zeta)\Leftrightarrow
 \Upsilon_n(z)\rightarrow e^{in\alpha}\Upsilon_n(z)\,, 
\end{eqnarray}
which can be regarded as chiral rotations of the 
 fermionic coordinates of the ${\cal N}=2$ superspace 
 (a diagonal group of automorphism group $SU(2)_R$).
This $U(1)$ action is precisely 
the one in Refs. \cite{Kaledin,Feix} 
acting on fiber.

In Ref. \cite{kuzenko}, it was claimed that 
 there exists a minimal extension of every four dimensional
 ${\cal N}=1$ supersymmetric nonlinear sigma model described 
 by (\ref{kahler1}) to four dimensional ${\cal N}=2$
 supersymmetric nonlinear sigma model described by (\ref{action2}).
Indeed,
 it is easy to see that the action (\ref{action2}) involves
 ${\cal N}=1$ K{\"a}hler potential and can be regarded as an ${\cal
 N}=2$ extension.
Representing $\Upsilon(\zeta)$ in the form
 $\Upsilon=\Phi+\zeta\Sigma+{\cal A}(\zeta)$ where ${\cal A}(\zeta)$ 
 contains all the
 auxiliary superfields, the action (\ref{action2}) can be 
 rewritten as
\begin{eqnarray}
 S = \int d^8z\left\{
 {1 \over 2\pi i}\oint {d\zeta \over \zeta}
 \exp{{\Bigg (}{\cal A}{\partial \over \partial \Phi}
 +{\bar{\cal A}}{\partial \over \partial \bar{\Phi}}{\Bigg )}}
 \exp\left[\zeta\Sigma{\partial \over \partial \Phi}
 -{1 \over \zeta}\bar{\Sigma}{\partial \over \partial \bar{\Phi}}\right]
 K(\Phi,\bar{\Phi})
\right\}\,.\label{action3}
\end{eqnarray}
One can see that an extension of ${\cal N}=1$ model into
 ${\cal N}=2$ nonlinear sigma model can be obtained
 via (\ref{action3}) with corresponding 
 ${\cal N}=1$ K{\"a}hler potential. 
However, since this action still includes the infinite tower of
 auxiliary superfields ${\cal A}$,
 we have to eliminate them by their equations of motion, 
 in order to obtain the action 
 in terms of physical superfields $\Phi$ and $\Sigma$ only. 
Their equations of motion read 
\begin{eqnarray}
 {1 \over 2\pi i}\oint {d\zeta \over \zeta}\zeta^n
 {\partial \over \partial\Upsilon^i_*}
 K(\Upsilon_*,\bar{\Upsilon}_*)=0\,,~~~~n\ge 2\, \label{aux}
\end{eqnarray} 
with $\Upsilon_*(\zeta)$ denoting a solution.

In general, it is difficult to solve (\ref{aux}) exactly and 
 the auxiliary fields can be eliminated at most perturbatively \cite{kuzenko}.
However, it was claimed in Ref. \cite{kuzenko}
that one can exactly solve Eq. (\ref{aux}) 
if the following conditions are satisfied:
\begin{itemize}
\item The K{\"a}hler manifold is a homogeneous
 space, a coset space $G/H$,  
 with an isometry $G$.
\item The K{\"a}hler potential is invariant 
 under the ${\cal N}=1$ $U(1)_R$ symmetry, defined by 
 $\Phi \to e^{i \alpha} \Phi$ and  $\Sigma \to e^{i \alpha} \Sigma$. 
\end{itemize}
The authors in Ref. \cite{kuzenko} 
 showed how to solve the equations (\ref{aux}) for 
 the ${\bf C}P^1$ base manifold explicitly and
 constructed ${\cal N}=2$ supersymmetric 
 nonlinear sigma models on the tangent $T{\bf C}P^{1}$ 
 and the cotangent $T^*{\bf C}P^1$ 
 bundles over ${\bf C}P^1$.
They also wrote down 
 the $T{\bf C}P^{n-1}$ model in Ref. \cite{kuzenko2}.
%without giving any explicit solution for $\Upsilon_*$.
In the following, we give a comprehensive
 review of how to obtain the nonlinear sigma
 model with $T^*{\bf C}P^{n-1}$. 
%with explicit solution $\Upsilon_*$.
Then, we consider the action with the tangent $T Q^{n-2}$ and
 cotangent $T^*Q^{n-2}$ bundles
 over quadric surface $Q^{n-2} = SO(n)/[SO(n-2)\times U(1)]$.

%%%%%%%%%%%%%%%%%%%%%%%%%%%%%%%%%%%%%%%%%%%%%%%%%%%%%%%%%%%%
\section{(Co)tangent bundle over ${{\bf C}P^{n-1}}$}
\begin{figure}
\begin{center}
  \epsfxsize=8cm
  \epsfbox{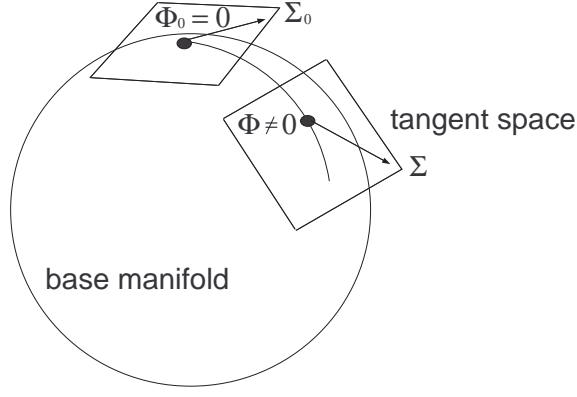}
\caption{Base manifold ${\bf C}P^{n-1}$ with its tangent spaces 
 at $\Phi=0$ and $\Phi\neq 0$. 
 One can arrive at any point $\Phi\neq 0$ from $\Phi_0=0$ 
 by $SU(n)$ transformation with parameters $\epsilon$ 
 and $\bar{\epsilon}$ on the manifold.
 By using the transformation law for $\Phi$,
 the tangent vector at $\Phi\neq 0$ is also obtained.}
\label{trans1}
\end{center}
\end{figure}

The K{\"a}hler potential of ${{\bf C}P^{n-1}}$ nonlinear sigma model
 is given by \footnote{We take the notation that
 repeated indices mean the summation with respect to them even if they
 are at the same upper(down) positions. For instance,
 $A^{ij}B^j\equiv \sum_jA^{ij}B^j$. Throughout this paper, 
 we respect
 upper and down indices as vector and covector, respectively.}
\begin{eqnarray}
 K(\Phi^i,\bar{\Phi}^{\bar j})=r^2 \ln
  \left(1+{|\Phi^i|^2 \over r^2}\right) \label{kahler_CP}
\end{eqnarray}
where $\Phi^i~(i=1,\cdots,n-1)$ are chiral superfields 
 and $r$ is a real constant with mass dimension one.
The potential for its ${\cal N}=2$ extension can be 
 obtained by replacing $\Phi^i$ in (\ref{kahler_CP})
 by the superfield $\Upsilon^i$.
The equations of motion to the infinite auxiliary 
 fields read from (\ref{aux}):
\begin{eqnarray}
 {1 \over 2\pi i}\oint {d\zeta \over \zeta}\zeta^m
 {r^2 \bar{\Upsilon}_*^{\bar{i}} 
 \over r^2 + |\Upsilon_*^j|^2}=0\,,~~~~~~m\ge 2\,. \label{eq_aux}
\end{eqnarray}
It is difficult to find a solution of Eq.~(\ref{eq_aux})
 at arbitrary 
 point of the base manifold ${\bf C}P^{n-1}$. 
However, one can readily 
 find the solution at the origin $\Phi=0$ as
\begin{eqnarray}
 \Upsilon_0^i=\Sigma_0^i \zeta,~~
 \bar{\Upsilon}_{0}^{\bar{i}}=-{\bar{\Sigma}_{0}^{\bar{i}} \over \zeta}\,.
 \label{solution}
\end{eqnarray}
Here $\Sigma_0^i$ is a tangent vector at the origin of $\Phi$.
We need a solution $\Upsilon^i_*$ at $\Phi\neq 0$.
In order to obtain it, 
 one should take into account that Eq. (\ref{eq_aux}) is invariant under the
 $SU(n)$ transformation because 
 ${\bf C}P^{n-1}$ is a homogeneous space with $SU(n)$ isometry 
and the
 K{\"a}hler potential (\ref{kahler_CP}) is invariant under $SU(n)$
 transformation up to a K\"ahler transformation. 
Then, applying $SU(n)$ transformation to the curve $\Upsilon_0^i(\zeta)$, 
 one can obtain the solution $\Upsilon^i_*(\zeta)$
 at $\Phi\neq 0$ (see Fig. \ref{trans1}).
%%%%%%%%%%%%%%%%%%%%%%%%%%%%%%%
%%%%%%%%%%%%%%%%%%%%%%%%%%%%%%

Let $\phi^i=(x,y^j)^T~(j=1,\cdots,n-1)$ be homogeneous  
 coordinates of ${\bf C}P^{n-1}$. 
 Infinistimal $SU(n)$ transformations for $\phi^i$ can be decomposed into 
 \cite{Higashijima:1999ki}
\begin{eqnarray}
 \delta \phi^i
 &=&(i\theta T + i\theta^A T_A+\bar{\epsilon}^{\bar{j}} E^j
   + \epsilon^j \bar{E}^{\bar{j}})\phi^i \nonumber \\
 &=&\left(
  \begin{array}{cc}
   i\sqrt{2(n-1) \over n}\theta & \bar{\epsilon}^{\bar{j}} \\
   -\epsilon^i & -i\theta^A\rho(T_A)^{ij}
   -i\sqrt{2 \over n(n-1)}\theta\delta^{ij}
   \end{array}
 \right) 
 \left(
  \begin{array}{c}
   x \\
   y^j
  \end{array}
 \right)\,, \label{trans_CP}
\end{eqnarray}
where 
 $T$ is the $U(1)$ generator, $E^i(\bar{E}^{\bar{i}}=(E^i)^\dagger)$ is 
 $n-1$ raising (lowering) operators represented by upper (lower) 
 triangle matrices, and $\rho(T_A)$ is an $n-1$ by $n-1$ matrix for
 the fundamental representation of $SU(n-1)$.
From Eq.
(\ref{trans_CP}), one can obtain 
 the action of the finite transformation on the homogeneous coordinate
\begin{eqnarray}
 \phi^{i} = \pmatrix { x \cr y^j}\rightarrow 
        \left(
        \begin{array}{cc}
          x\cos A + {\bar{\epsilon}\cdot y \sin A \over A} \\
          -{\epsilon^j x\sin A \over A}+y^j
          -{\epsilon^j\bar{\epsilon}\cdot y \over A^2}(1-\cos A)
        \end{array}
        \right)\,, \label{finite}
\end{eqnarray}
with $A\equiv\epsilon^i\bar{\epsilon}_{\bar{i}}=\epsilon\cdot \bar{\epsilon}$. 
Here we take $\theta=\theta^A=0$ since such a parameter 
 choice is sufficient to 
 consider the point $\Phi\neq 0$.
Using Eq. (\ref{finite}), one obtains the transformation law for
 inhomogeneous coordinates $\Phi^i \equiv ry^i/x$ ($i=1,\cdots,n-1$)
 of the ${\bf C}P^{n-1}$ as 
\begin{eqnarray}
 \Phi^{i}
  =f^i(\Phi_0^j)
  ={{-r\epsilon^i\sin A \over A \cos A}+{\Phi_0^i \over \cos A}
   -{\epsilon^i(\bar{\epsilon} \cdot \Phi_0) \over A^2 \cos A}(1-\cos A)
   \over 1 + {\bar{\epsilon}\cdot \Phi_0 \sin A \over rA \cos A}}\,.
  \label{trans2}
\end{eqnarray}
Eq. (\ref{trans2}) tells us
 a transformation law for the tangent vector $\Sigma^i$
\begin{eqnarray}
\Sigma^{i}=
 {\partial f^i(\Phi) \over \partial \Phi^j}{\Bigg |}_{\Phi=\Phi_0}\Sigma_0^j\,.
 \label{trans_sigma}
\end{eqnarray}
Substituting $\Phi_0=0$ into (\ref{trans2}) and (\ref{trans_sigma}), one
 obtains the transformation law from the point
 $\Phi_0=0$ to $\Phi\neq 0$ as
\begin{eqnarray}
 \Phi^{i}&=&f^i(\Phi_0=0)=-{r\epsilon^i \sin A \over A\cos A}\,, \label{s1} \\
 \Sigma^{i}&=&{\partial f^i(\Phi) \over \partial \Phi^j}{\Bigg |}_{\Phi=0}
           \Sigma_0^j=
 \left\{{\delta^{ij} \over \cos A}
 -{\epsilon^i\bar{\epsilon}^{\bar{j}} \over A^2 \cos^2 A}(\cos A-1)\right\}
 \Sigma_0^j\equiv V^i_{~j}(\epsilon,\bar{\epsilon})\Sigma_0^j\,. \label{s3}
\end{eqnarray} 
Here $V^{i}_{~j}$ can be written in terms of $\Phi^i$ with (\ref{s1}) as
\begin{eqnarray}
  V^{i}_{~j}
 &=&\sqrt{1+{|\Phi^k|^2 \over r^2}}\left(\delta^{ij}
 -{\Phi^i\bar{\Phi}^{\bar{j}} \over 
 |\Phi^l|^2}\left(1-\sqrt{1+{|\Phi^m|^2 \over r^2}}
  \right)\right)\,,
\end{eqnarray}
and its inverse is given by
\begin{eqnarray}
 (V^{-1})^{i}_{~j}={1 \over \sqrt{1+|\Phi^k|^2 \over r^2}}
               \left\{\delta^{ij}-{\Phi^i\bar{\Phi}^{\bar{j}}\over |\Phi^l|^2}
               \left(1-{1 \over \sqrt{1+|\Phi^m|^2 \over r^2}}\right)
     \right\}\,. \label{s2}
\end{eqnarray}

Taking that $\Upsilon^i$ follows the same transformation 
 law with (\ref{trans2}) into account, 
 one finds
\begin{eqnarray}
  \Upsilon_*^i&=&f^i(\Upsilon_0)
 ={{-r\epsilon^i\sin A \over A \cos A}+{\Upsilon_0^i \over \cos A}
   -{\epsilon^i(\bar{\epsilon} \cdot \Upsilon_0) \over A^2 \cos A}(1-\cos A)
   \over 1 + {\bar{\epsilon}\cdot \Upsilon_0 \sin A \over rA \cos A}}.
   \label{sol_upsilon}
\end{eqnarray}
By using (\ref{solution}) and (\ref{s1}), Eq. (\ref{sol_upsilon}) can be
 rewritten in terms of $\Phi^i$ and $\Sigma_0^i$ as
\begin{eqnarray}
 \Upsilon_*^i={\Phi^i+D^{i}_{~j}\Sigma_0^j \zeta \over
             1-(\bar{\Phi}\cdot\Sigma_0)\zeta/r^2}\,, 
\end{eqnarray}
with 
\begin{eqnarray}
 D^{i}_{~j} \equiv 
  \sqrt{1+{|\Phi^k|^2 \over r^2}}
          \left(\delta^{ij}-{\Phi^i\bar{\Phi}^{\bar{j}} \over |\Phi^l|^2}\right)
          +{\Phi^i\bar{\Phi}^{\bar{j}} \over |\Phi^k|^2}\,. \label{trans}
\end{eqnarray}
Using (\ref{s3}) and (\ref{s2}), we obtain a 
 simple form of the solution
\begin{eqnarray}
 \Upsilon_*^i={\Phi^i(1+|\Phi^j|^2/r^2)
  +\zeta\left\{\Sigma^i(1+|\Phi^j|^2/r^2)
  -\Phi^i(\bar{\Phi}\cdot\Sigma)/r^2 \right\} \over
  1+|\Phi^k|^2/r^2-\zeta(\bar{\Phi}\cdot\Sigma)/r^2}\,.\label{upsilon-CP}
\end{eqnarray} 
It can be easily checked that this is actually the solution 
 of (\ref{eq_aux}) by
 substituting (\ref{upsilon-CP}) into (\ref{eq_aux}).
%In Ref. \cite{kuzenko2}
% the explicit solution was not given though
% the final tangent bundle action (see (\ref{final-CP}))
% was written down \cite{kuzenko2}.
 
Replacing $\Phi^i$ in the K\"ahler potential (\ref{kahler_CP}) 
by $\Upsilon_*^i$ in Eq.~(\ref{upsilon-CP}), 
we obtain 
\begin{eqnarray}
 K=\Psi+\bar{\Psi}
 +r^2\ln\left(1+{|\Phi^i|^2 \over r^2}\right)
 +r^2\ln\left\{1-{1 \over r^2\left(1+{|\Phi^j|^2 \over r^2}\right)}
         \left(|\Sigma^i|^2+{|\Sigma\cdot \bar{\Phi}|^2 \over r^2
               \left(1+{|\Phi^i|^2 \over r^2}\right)}\right)\right\}\,,
 \label{final-CP}
\end{eqnarray} 
with $\Psi$ defined by 
\begin{eqnarray}
 \Psi&\equiv& -r^2\ln\left(1-{\zeta \over r^2}
   {\bar{\Phi}\cdot\Sigma \over 1+|\Phi^i|^2/r^2} \right)\,.
\end{eqnarray}
After integrating over 
 the projective coordinate and
 introducing
 the K{\"a}hler metric of ${\bf C}P^{n-1}$ 
\begin{eqnarray}
g_{i\bar{j}}={r^2 \delta_{i\bar{j}} \over r^2+|\Phi^k|^2} 
 -{r^2 \bar{\Phi}^{\bar{i}}\Phi^j \over (r^2+|\Phi^k|^2)^2}\,, \label{kahler-cp}
\end{eqnarray}
we finally obtain the following action 
\begin{eqnarray}
 S=\int d^8 z 
     \left\{
       r^2\ln\left(1+{|\Phi^i|^2 \over r^2}\right)
         +r^2\ln
         \left(
          1-{1\over r^2}g_{i\bar{j}}\Sigma^i\bar{\Sigma}^{\bar{j}}
         \right)
     \right\}. \label{final_CP}
\end{eqnarray} 
This is the action of the nonlinear sigma model with the tangent bundle 
$T{\bf C}P^{n-1}$. 
Since the second term vanishes when $\Sigma =0$ holds,  
the first term is the K{\"a}hler potential of the base ${\bf C}P^{n-1}$ and the
 second term represents the tangent space. 
This form was 
 obtained for $T {\bf C}P^1$ 
 in Ref. \cite{kuzenko}
 and
 for $T {\bf C}P^{n-1}$ in Ref. \cite{kuzenko2}.

In order to obtain the cotangent bundle, 
 we need dualize the complex linear superfields $\Sigma^i$ in the action (\ref{final_CP}) 
 into chiral superfields $\psi_i$ being components of a cotangent vector 
 via the Legendre transformation. 
The resultant action can be identified with a hyper-K{\"a}hler
 potential.
The action can be dualized if we replace the action (\ref{final_CP}) by
 the following one:
\begin{eqnarray}
  S=\int d^8 z 
     \left\{
       r^2\ln\left(1+{|\Phi^i|^2 \over r^2}\right)
         +r^2\ln
         \left(
          1-{1\over r^2}g_{i\bar{j}}U^i\bar{U}^{\bar{j}}
         \right)
       +U^i\psi_i+\bar{U}^{\bar{i}}\bar{\psi}_i
     \right\}\,, \label{tan-led}
\end{eqnarray}
where $U^i$ is the complex unconstrained auxiliary superfields.
By the construction, $U^i$ is a tangent vector at point $\Phi$ of the
 manifold and therefore $\psi_i$ is a one-form at the same point.
Eliminating the auxiliary variables $U^i$ and $\bar{U}^{\bar{j}}$, with
 the aid of the equations of motion
\begin{eqnarray}
0&=&{\partial S \over \partial U^i}
 ={-g_{i\bar{j}}\bar{U}^{\bar{j}} 
  \over 1-g_{i\bar{j}}U^i\bar{U}^j/r^2}
  +\psi_i\,, \label{eq_cotangent1} \\
0&=&{\partial S \over \partial \bar{U}^i}
 ={-g_{j\bar{i}}U^j \over 1-g_{i\bar{j}}U^i\bar{U}^j/r^2}
  +\bar{\psi}_{\bar{i}}\, \label{eq_cotangent2}
\end{eqnarray}
 results in a purely chiral sigma model which is dually equivalent to
 the ${\cal N}=2$ supersymmetric model (\ref{final_CP}) and is defined on
 the cotangent bundle $T^*{\bf C}P^{n-1}$.
The final result is \cite{Curtright:1979yz,perelomov}
\begin{eqnarray}
 S=\int d^8 z \left\{r^2\ln\left(1+{|\Phi^i|^2 \over r^2}\right)
  -r^2\ln\left(f(\kappa)
   \right)+2r^2{\kappa \over f(\kappa)}\right\}\,,\,\,\,\,
 f(\kappa)={1\over 2}(1+\sqrt{1+4\kappa})\,,\label{cot-cp}
\end{eqnarray}
where $\kappa \equiv g^{i\bar{j}}\psi_i\bar{\psi}_{\bar{j}}/r^2$
 and $g^{i\bar{j}}$ the inverse metric of $g_{i\bar{j}}$.

Here we solved the equation of motion (\ref{eq_aux}) to remove the
 auxiliary fields.
There is an another method to eliminate the infinite auxiliary fields,
 which is briefly reviewed in Appendix A.

%%%%%%%%%%%%%%%%%%%%%%%%%%%%%%%%%%%%%%%%%%%%%%%%%%%%%%%%%%%%%%%%%%%%%%%%%%
\section{Construction of tangent bundle $TQ^{n-2}$}

In this section we construct the nonlinear sigma model with
 tangent bundle $TQ^{n-2}\simeq T[{SO(n) \over SO(n-2)\times U(1)}]$.
Here we use a strategy different from the last section, 
because it is difficult to solve the equations of motion (\ref{eq_aux}) 
for elimination of auxiliary fields in this case. 
First we consider the $n=4$ case of $Q^2$ 
(called the Klein quadric) by noting 
the isomorphism $Q^2\simeq SO(4)/[SO(2)\times U(1)]\simeq 
 [SU(2)\times SU(2)]/[U(1)\times U(1)]\simeq {\bf C}P^1\times {\bf C}P^1$.
We then solve the
 equations of motion for auxiliary fields 
in this $T Q^2$ case.  
We can show that the solution can be 
extended to the $TQ^{n-2}$ case, 
and then construct the $TQ^{n-2}$ model.

%%%%%%%%%%%%%%%%%%%%%%%%%%%%%%%%%%%%%%%%%%%%%%%%%%%%%%%%%%%%%%
\subsection{First look: $T Q^2$}\label{TCP1}
Let us start in the base manifold.
Considering the isomorphism mentioned above, 
 the K{\"a}hler potential of the nonlinear sigma model with $Q^2$ can
 be described by
\begin{eqnarray}
 K(\Phi^i,\bar{\Phi}^j)
  &=&r^2\ln\left(1+{|\Phi^1|^2 \over r^2}\right)
    +r^2\ln\left(1+{|\Phi^2|^2 \over r^2}\right) \nonumber \\
  &=&r^2\ln\left(1+{|\Phi^i|^2 \over r^2}
    +{|\Phi^1|^2|\Phi^2|^2 \over r^4}\right)\,,
  ~~~~~i=1,2\,. \label{q2}
\end{eqnarray}
Using the unitary transformation
\begin{eqnarray}
 \left(
 \begin{array}{c}
  \Phi^1 \\
  \Phi^2
 \end{array}
 \right)\rightarrow
 {1 \over \sqrt{2}}\left(\begin{array}{cc}
		    1 & -i \\
                    1 & i
			 \end{array}
                    \right)
 \left(
 \begin{array}{c}
  \Phi^1 \\
  \Phi^2
 \end{array}
 \right)\,,\label{unitary}
\end{eqnarray}  
 we arrive at the K\"ahler potential (\ref{kahler_SO})  with $i=1,2$ 
for $Q^2$. 
The K{\"a}hler metric is then given by Eq.~(\ref{metric_quad}) with $i=1,2$.

Next we consider its tangent bundle.
Replacing the chiral superfield $\Phi^i$ in Eq.~(\ref{q2}) 
by the scalar multiplet $\Upsilon^i$, 
 we arrive at the action of the tangent bundle: 
\begin{eqnarray}
 S&=&\int {d^8z}{1 \over 2\pi i}\oint {d\zeta \over \zeta}
   \left\{K(\Upsilon^1,\bar{\Upsilon}^1)
    + K(\Upsilon^2,\bar{\Upsilon}^2)\right\} \label{cp-cp}
\end{eqnarray}
with  
\begin{eqnarray}
 K(\Upsilon^i,\bar{\Upsilon}^i)
 =r^2\ln\left(1+{|\Upsilon^i|^2 \over r^2}\right)\,, 
 ~~~~~\mbox{(no sum)}\,.\label{cp1}
\end{eqnarray}
Since the calculation for the ${\bf C}P^{1}$ case 
 in the previous section can be done 
 independently for each term in (\ref{cp-cp}),
 the resultant action is just a sum of the action (\ref{final_CP})
 with a different kind of chiral and complex linear superfields:
\begin{eqnarray}
 S&=&\int d^8z 
  \left\{r^2\ln\left(1+{|\Phi^i|^2 \over r^2}
   +{|\Phi^1|^2|\Phi^2|^2 \over r^4}\right)
   \right. \nonumber \\
 && 
  \left.
  +r^2\ln\left(1-\sum_{i=1}^2G_i{|\Sigma^i|^2 \over r^2}
  +{1 \over r^4}G_1G_2|\Sigma^1|^2|\Sigma^2|^2 \right)\right\}\,,
  \label{action-base}
\end{eqnarray}
where $G_i$ is the metric of ${\bf C}P^1$ given by
\begin{eqnarray}
 G_i={1 \over \left(1+{|\Phi^i|^2 \over r^2}\right)^2}\,,
 ~~~~~\mbox{(no sum)}\,.\label{metriccp}
\end{eqnarray}
In Eq. (\ref{action-base}), the first term is 
 the K{\"a}hler potential of the base manifold $Q^2$
 and the second one is of tangent one.
This form of the action is in particular coordinates 
and so it is better to  
 rewrite it by geometrical quantities such as 
 K{\"a}hler potential, metric and Riemann tensor.
Indeed, as shown in (\ref{kahler_SO}),
 the first term is written by K{\"a}hler potential
 after performing unitarity transformation.
Let us focus on the tangent vector sector (second logarithm).
Taking that
 the tangent vector $\Sigma^i$ follows the same unitary transformation
 law with one of $\Phi^i$ into account, 
 we find that the term $\sum_{i=1}^2G_i|\Sigma^i|^2/r^2$
 is the form of 
 $g_{i{\bar{j}}}\Sigma^i\bar{\Sigma}^{\bar{j}}$, 
 where $g_{i\bar{j}}$ is
 a metric of quadric surface given in Eq.~(\ref{metric_quad}).
The last term in the second logarithm, 
 $G_1G_2|\Sigma^1|^2|\Sigma^2|^2$, can be rewritten by 
 the covariant quantity by noting that it is in forth order 
in $\Sigma$ or $\bar \Sigma$. 
First, we calculate $(g_{i{\bar{j}}}\Sigma^i\bar{\Sigma}^{\bar{j}})^2$
 in the frame taken in (\ref{action-base}).
\begin{eqnarray}
 (g_{i{\bar{j}}}\Sigma^i\bar{\Sigma}^{\bar{j}})^2
 =G_1^2|\Sigma^1|^4+G_2^2|\Sigma^2|^4+2G_1G_2|\Sigma^1|^2|\Sigma^2|^2\,. 
 \label{a1}
\end{eqnarray}  
Another forth order term is 
\begin{eqnarray}
 R_{i\bar{j}k\bar{l}}
\Sigma^i\bar{\Sigma}^{\bar{j}}\Sigma^k\bar{\Sigma}^{\bar{l}} 
 &=&-{2|\Sigma^1|^4 \over r^2\left(1+{|\Phi^1|^2 \over r^2}\right)^4} 
 -{2|\Sigma^2|^4 \over r^2\left(1+{|\Phi^2|^2 \over r^2}\right)^4} \nonumber \\
 &=&-{2 \over r^2}(G_1^2|\Sigma^1|^4+G_2^2|\Sigma^2|^4)\,,\label{a2}
\end{eqnarray}
where $R_{i\bar{j}k\bar{l}}$ is Riemann tensor defined by 
 $R_{i\bar{j}k\bar{l}}=\partial_{k}\partial_{\bar{l}}g_{i\bar{j}}
 -g^{m\bar{n}}\partial_mg_{i\bar{j}}\partial_{\bar{n}}g_{k\bar{l}}$.
Using (\ref{a1}) and (\ref{a2}), we get the relation
\begin{eqnarray}
 G_1G_2|\Sigma^1|^2|\Sigma^2|^2
 ={1 \over 2}\left\{
 (g_{i{\bar{j}}}\Sigma^i\bar{\Sigma}^{\bar{j}})^2+{r^2 \over 2}
  R_{i\bar{j}k\bar{l}}
 \Sigma^i\bar{\Sigma}^{\bar{j}}\Sigma^k\bar{\Sigma}^{\bar{l}} 
 \right\}\,.\label{a3}
\end{eqnarray}
Putting all together, and performing the unitary
 transformation, 
 we finally obtain the action of the tangent bundle $TQ^{2}$
 over quadric surface $Q^2$
\begin{eqnarray}
  S&=&\int d^8z 
  \left[r^2\ln\left(1+{|\Phi^i|^2 \over r^2}
   +{(\Phi^i)^2(\bar{\Phi}^{\bar{j}})^2 \over 4r^4}\right)
   \right. \nonumber \\
 && 
  \left.
  +r^2\ln\left\{1-{1 \over r^2} g_{i\bar{j}}
    \Sigma^i\bar{\Sigma}^{\bar{j}}
  + {1 \over 2r^4}\left((g_{i\bar{j}}
    \Sigma^i\bar{\Sigma}^{\bar{j}})^2+{r^2 \over 2}
    R_{i\bar{j}k\bar{l}}\Sigma^i\bar{\Sigma}^{\bar{j}}
    \Sigma^k\bar{\Sigma}^{\bar{l}}\right)\right\}\right]\,. \label{action-tan}
\end{eqnarray}
There appear the fourth order terms of the tangent vector.
This is a particular form of the general expression 
of the hyper-K\"ahler sigma models 
suggested in (2.47) in Ref. \cite{kuzenko}.
The method which we use 
 is simpler and easier than solving the equation of
 motion (\ref{eq_aux}) as performed in the previous section,
 in which one may suffer from 
 rewriting the transformation parameters $\epsilon^i$ 
 and $\bar{\epsilon}^{\bar{i}}$ in terms of $\Phi^i$ as has been done
 in (\ref{upsilon-CP})
 and also from finding the form of 
the tangent bundle action (\ref{action-tan}).
In the following subsection, we solve the equations of motion for
 auxiliary fields for $TQ^{n-2}$ case
 as the same with the $T{\bf C}P^{n-1}$ case.
Then, we show that the tangent bundle action (\ref{action-tan})
 is also valid in $TQ^{n-2}$ with simply extending 
the index $i$ to run from $1$ to $n-2$.

%%%%
\subsection{Solving the equations of motion and deriving 
 $TQ^{n-2}$ action}
We start with ${\cal N}=2$ extension of the K{\"a}hler potential of 
 $Q^{n-2}$
\begin{eqnarray}
K(\Upsilon^i,\bar{\Upsilon}^{\bar{j}})=
 r^2\ln\left(1+{|\Upsilon^i|^2 \over r^2}
 +{(\Upsilon^i)^2(\bar{\Upsilon}^{\bar{j}})^2 \over 4 r^4}\right)\,.
 \label{kahler_SOn}
\end{eqnarray}
The equation of motion for auxiliary fields of the polar multiplet reads
\begin{eqnarray}
 {1 \over 2\pi i}\oint_C {d\zeta \over \zeta}\zeta^m
 {r^2 \bar{\Upsilon}_*^i+{1\over 2}
 \Upsilon_*^i(\bar{\Upsilon}_*^{\bar{j}})^2 \over r^4+r^2|\Upsilon_*^k|^2
 +{1 \over 4}(\Upsilon_*^k)^2(\bar{\Upsilon}_*^{\bar{l}})^2}=0\,,~~~~m\ge 2\,.
 \label{eq_auxiliary2}
\end{eqnarray}
It is easy to check that the same solution with (\ref{solution}) 
 also satisfies this equation.
We can obtain the solution at $\Phi\neq 0$
 by applying the finite
 $SO(n)$ transformation to the curve (\ref{solution})
 since (\ref{eq_auxiliary2}) is $SO(n)$ invariant.

In the following we again focus on the $n=4$ case, $TQ^{2}$, for a while.
Using Eq. (\ref{trans_SO}) leads to the 
 finite $SO(4)$ transformation law for the homogeneous 
 coordinates as
\begin{eqnarray}
x&=&{1 \over 2}(\cos\beta_++\cos\beta_-)x_0
 +{\beta_+(\bar{\gamma}_+\cdot y_0)\sin\beta_-
  +\beta_-(\bar{\gamma}_-\cdot y_0)\sin\beta_+ 
   \over 2 \lambda \beta_+ \beta_-} \nonumber \\
 &&-{\sqrt{(\bar{\epsilon}^{\bar{i}})^2}(\cos\beta_+-\cos\beta_-) 
   \over 2 \sqrt{(\epsilon^i)^2}}z_0\,, \label{trans-so} \\
y^i
 &=&{1 \over 2\lambda}\left[
  (\cos\beta_--\cos\beta_+)\{\epsilon^i(\bar{\epsilon}\cdot y_0)
   -y_0^i(\epsilon \cdot \bar{\epsilon})+\bar{\epsilon}^{\bar{i}}
    (\epsilon\cdot y_0)\}
   +\lambda(\cos\beta_++\cos\beta_-)y_0^i
 \right] \nonumber \\
 &&-{\beta_+\gamma_+^i\sin\beta_-+\beta_-\gamma_-^i\sin\beta_+ \over
  2\lambda \beta_+\beta_-}x_0
 -{\beta_+\bar{\gamma}_{+}^{\bar{i}}\sin\beta_-
 +\beta_-\bar{\gamma}_{-}^{\bar{i}}\sin\beta_+ \over 2\lambda \beta_+\beta_-}z_0\,,
 \label{trans-so2}
\end{eqnarray}
where
\begin{eqnarray}
\lambda &\equiv& \sqrt{(\epsilon^i)^2}\sqrt{(\bar{\epsilon}^{\bar{j}})^2}\,, 
  \label{def1}\\
\beta_+ &\equiv &\sqrt{\epsilon\cdot \bar{\epsilon}-\lambda}\,,~~~~~ 
\beta_- \equiv \sqrt{\epsilon \cdot \bar{\epsilon}+\lambda}, \label{def2} \\
\gamma_+^i &\equiv & (\epsilon^j)^2\bar{\epsilon}^{\bar{i}}+\epsilon^i\lambda\,,~~~~~
\gamma_-^i \equiv -(\epsilon^j)^2\bar{\epsilon}^{\bar{i}}+\epsilon^i\lambda\,.
 \label{def3}
\end{eqnarray}
Using the definition of the inhomogeneous coordinates of quadric surface
  $\Phi^i \equiv ry^i/x$,
 the transformation laws for $\Phi^i$ and $\Sigma^i$ 
 are given by
\begin{eqnarray}
\Phi^{i}&=&r{y^{i} \over x}{\Bigg|}_{\Phi_0=0}
=-r{\beta_+\gamma_+^i\sin\beta_-+\beta_-\gamma_{-}^i\sin\beta_+ \over 
  \lambda \beta_+\beta_-(\cos\beta_++\cos\beta_-)}\,, \label{phi-so} \\
\Sigma^{i}&=&\left[{  
 (\cos\beta_--\cos\beta_+)(\epsilon^i\bar{\epsilon}^{\bar{j}}
 -\delta^{ij}(\epsilon\cdot\bar{\epsilon})
 +\bar{\epsilon}^{\bar{i}}\epsilon^j)
\over \lambda(\cos\beta_++\cos\beta_-)}
 +\delta^{ij}+{\Phi^{i}\bar{\Phi}^{\bar{j}} \over r^2} \right]\Sigma_0^j\, 
 \nonumber \\
&\equiv& (V_{Q})^{i}_{~j}\Sigma_0^j\,. \label{sigma-so}
\end{eqnarray}
Here, in the first line in Eq. (\ref{sigma-so}), 
 we have replaced the third term written 
 by $\epsilon$ and $\bar{\epsilon}$ 
 with $\Phi^{i}$ and $\bar{\Phi}^{\bar{j}}$ defined in (\ref{phi-so}). 
Since the transformation law for $\Upsilon^i$ is the same with 
 one of $\Phi^i$, we can obtain the $\Upsilon_*^i$ at non-zero value of
 $\Phi^i$ from (\ref{trans-so}) and (\ref{trans-so2}) as 
\begin{eqnarray}
\Upsilon_*^i&=&\left[{\cos\beta_+\{-\epsilon^i(\bar{\epsilon}\cdot\Upsilon_0)
 +\beta_-^2\Upsilon_0^i-\bar{\epsilon}^{\bar{i}}(\epsilon\cdot \Upsilon_0)\}
 +\cos\beta_-\{\epsilon^i(\bar{\epsilon}\cdot\Upsilon_0)-\beta_+^2\Upsilon_0^i
 +\bar{\epsilon}^{\bar{i}}(\epsilon\cdot\Upsilon_0)\} 
 \over \lambda(\cos\beta_++\cos\beta_-)}
\right. \nonumber \\
&&-r{\beta_+\gamma_+^i\sin\beta_-
 +\beta_-\gamma_-^i\sin\beta_+ \over \lambda\beta_+\beta_-(\cos\beta_++\cos\beta_-)}
 +{\beta_+\bar{\gamma}_{+}^{\bar{i}}
 \sin\beta_-+\beta_-\bar{\gamma}_{-}^{\bar{i}}\sin\beta_+ 
 \over 2r\lambda\beta_+\beta_-(\cos\beta_++\cos\beta_-)}(\Upsilon_0^l)^2{\Bigg ]} 
 \nonumber \\
&&\times\left[
1+{\beta_+(\bar{\gamma}_+\cdot\Upsilon_0)\sin\beta_-
   +\beta_-(\bar{\gamma}_-\cdot \Upsilon_0)\sin\beta_+ \over 
    r\lambda \beta_+\beta_-(\cos\beta_++\cos\beta_-)}
   +{\sqrt{(\bar{\epsilon}^{\bar{i}})^2}(\cos\beta_+-\cos\beta_-) \over
     2r^2\sqrt{(\epsilon^j)^2}(\cos\beta_++\cos\beta_-)}(\Upsilon_0^k)^2
\right]^{-1}\,.\nonumber \\ \label{upsilon1-so}
\end{eqnarray}

Let us rewrite Eq. (\ref{upsilon1-so}) in terms of $\Phi^i$ and $\Sigma^i$.
In order to do that, we need some formulas.
Using the expression (\ref{phi-so}) we have 
\begin{eqnarray}
 (\Phi^i)^2&=&{2r^2(\epsilon^i)^2(\cos\beta_+-\cos\beta_-) 
   \over \lambda(\cos\beta_++\cos\beta_-)}\,, \label{formula-so2} \\
 (\bar{\Phi}^{\bar{i}})^2&=&{2r^2(\bar{\epsilon}^{\bar{i}})^2
 (\cos\beta_+-\cos\beta_-) 
   \over \lambda(\cos\beta_++\cos\beta_-)}\,. \label{formula-so3}
\end{eqnarray}
Multiplying (\ref{formula-so2}) and (\ref{formula-so3}), we get
\begin{eqnarray}
 (\Phi^i)^2(\bar{\Phi}^{\bar{j}})^2={4r^4(\cos\beta_+-\cos\beta_-)^2 
  \over (\cos\beta_++\cos\beta_-)^2}. \label{formula-so4}
\end{eqnarray}
Dividing (\ref{formula-so2}) by (\ref{formula-so3}), we find
\begin{eqnarray}
 {\sqrt{(\bar{\epsilon}^{\bar{i}})^2} \over \sqrt{(\epsilon^j)^2}}
 ={\sqrt{(\bar{\Phi}^{\bar{i}})^2} \over \sqrt{(\Phi^j)^2}}\,. \label{formula-so5}
\end{eqnarray} 

Using (\ref{phi-so}), (\ref{formula-so4})
 and (\ref{formula-so5}), Eq. (\ref{upsilon1-so}) can be rewritten as
\begin{eqnarray}
\Upsilon*^i&=&\left[{\cos\beta_+\{-\epsilon^i(\bar{\epsilon}\cdot\Upsilon_0)
 +\beta_-^2\Upsilon_0^i-\bar{\epsilon}^{\bar{i}}(\epsilon\cdot \Upsilon_0)\}
 +\cos\beta_-\{\epsilon^i(\bar{\epsilon}\cdot\Upsilon_0)-\beta_+^2\Upsilon_0^i
 +\bar{\epsilon}^{\bar{i}}(\epsilon\cdot\Upsilon_0)\} \over 
 \lambda(\cos\beta_++\cos\beta_-)}
\right. \nonumber \\
&&+\Phi^i-{1\over 2r^2}\bar{\Phi}^{\bar{i}}(\Upsilon_0^j)^2 {\Bigg ]}
\left[
 1-{\bar{\Phi}\cdot \Upsilon_0 \over r^2}
 +{(\bar{\Phi}^{\bar{i}})^2 \over 4r^4}(\Upsilon_0^k)^2
\right]^{-1}\,. 
\end{eqnarray}
Further, using (\ref{sigma-so}) with $\Upsilon_0^i=\zeta\Sigma_0^i$, we 
 obtain a simple form
\begin{eqnarray}
 \Upsilon*^i={\Phi^i+\zeta\left(\Sigma^i
             -{\Phi^i(\bar{\Phi}(V_Q^{-1})\Sigma) \over r^2}\right)
             -\zeta^2{\bar{\Phi}^{\bar{i}}
              (\Sigma^t(V_Q^{-1})^tV_Q^{-1}\Sigma) \over 2r^2}
            \over 
             1-\zeta{(\bar{\Phi}V_Q^{-1}\Sigma)\over r^2}
             +\zeta^2{(\bar{\Phi}^{\bar{k}})^2
              (\Sigma^t(V_Q^{-1})^tV_Q^{-1}\Sigma)\over 4r^4}}\,.
 \label{upsilon3-so}
\end{eqnarray}
To complete the work, we have to find the expression of $V_Q^{ij}$
 in terms of $\Phi^i$.
The result is
\begin{eqnarray}
 V_Q&=&\left(
        \begin{array}{cc}
	 1+{|\Phi^k|^2 \over 2r^2} 
         & {\Phi^1\bar{\Phi}^{\bar{2}}-\bar{\Phi}^{\bar{1}}\Phi^2 \over 2r^2} \\
         -{\Phi^1\bar{\Phi}^{\bar{2}}-\bar{\Phi}^{\bar{1}}\Phi^2 \over 2r^2} 
         & 1+{|\Phi^k|^2 \over 2r^2} \\
	\end{array}
     \right)\,,\\
 V_Q^{-1}&=&L^{-1}
       \left(
        \begin{array}{cc}
	 1+{|\Phi^k|^2 \over 2r^2} 
         & -{\Phi^1\bar{\Phi}^{\bar{2}}-\bar{\Phi}^{\bar{1}}\Phi^2 \over 2r^2} \\
         {\Phi^1\bar{\Phi}^{\bar{2}}-\bar{\Phi}^{\bar{1}}\Phi^2 \over 2r^2} 
         & 1+{|\Phi^k|^2 \over 2r^2} \\
	\end{array}
       \right)\,,\label{t2}
\end{eqnarray}
with  
\begin{eqnarray}
 L\equiv 1+{|\Phi^i|^2 \over r^2}
 +{(\Phi^i)^2(\bar{\Phi}^{\bar{j}})^2 \over 4r^4}\,.
\end{eqnarray}
Eq. (\ref{t2}) leads 
 $(V_Q^{-1t}V_Q^{-1})^{ij}=L^{-1}\delta^{ij}$
 and we obtain the final form of the solution
\begin{eqnarray}
 \Upsilon^i_*={\Phi^i+\zeta\left\{\Sigma^i-{\Phi^i\over r^2L}
      \left(\bar{\Phi}\cdot\Sigma
     +{(\Phi\cdot\Sigma)(\bar{\Phi}^j)^2 \over 2r^2}\right)\right\}
     -\zeta^2{\bar{\Phi}^i(\Sigma^j)^2 \over 2r^2L} 
 \over 1-{\zeta \over r^2L}\left\{\bar{\Phi}\cdot\Sigma
     +{(\Phi\cdot\Sigma)(\bar{\Phi}^k)^2 \over 2r^2}\right\}
       +\zeta^2{(\bar{\Phi}^k)^2(\Sigma^l)^2 \over 4r^4L}}\,.
 \label{sol-upsilon}
\end{eqnarray}
This solution satisfies the equation of motion 
 (\ref{eq_auxiliary2}) for the $TQ^2$ case of $n=4$.

Let us turn back to general $n$. 
When one tries to derive the solution $\Upsilon_*^i$ 
for higher dimensional case $(n> 4)$, 
it might be problematic to convert
 the transformation parameters $\epsilon^i$ and $\bar{\epsilon}^{\bar{i}}$ in
 $V_Q$ and $V_Q^{-1}$ into the base coordinates $\Phi$ and eventually
 to find the final form of the solution $\Upsilon_*$.
But fortunately 
if one respects the solution (\ref{sol_upsilon}) of $n=4$ 
by making the index $i$ run from $1$ to $n-2$, 
it becomes a solution for general $n$.  
Indeed, one can straightforwardly check that 
it satisfies the equations of motion (\ref{eq_auxiliary2})
 by substituting Eq.~(\ref{sol-upsilon}) into (\ref{eq_auxiliary2}):
\begin{eqnarray}
&&{1 \over 2\pi i}\oint_C {d\zeta \over \zeta}\zeta^m
 {r^2 \bar{\Upsilon}_*^i+{1\over 2}
 \Upsilon_*^i(\bar{\Upsilon}_*^{\bar{j}})^2 \over r^4+r^2|\Upsilon_*^k|^2
 +{1 \over 4}(\Upsilon_*^k)^2(\bar{\Upsilon}_*^{\bar{l}})^2} \nonumber \\
&&={1 \over 2\pi i}
 \oint_C {d\zeta \over \zeta}\zeta^{m-2}{8r^4 \over g(\Phi,\Sigma)}
 \left[4r^4(\bar{\Sigma^i})^2 Y^i-8r^2\zeta(\bar{\Phi}\cdot\bar{\Sigma})
  (r^2+|\Phi^j|^2)Y^i +\zeta^2(\Phi^j)^2(\bar{\Phi}^k)^4Y^i \right. \nonumber \\
&&~+8r^4\zeta^2(r^2+|\Phi^j|^2-\zeta(\bar{\Phi}\cdot\Sigma))\bar{Y}^i
  +2(\bar{\Phi^j})^2r^2\zeta\left\{2(\Phi\cdot\bar{\Sigma})Y^i+\zeta(2|\Phi^j|^2Y^i
   +2r^2Y^i+(\Phi^j)^2\bar{Y}^i \right.\nonumber \\
&&~\left. \left.-2\zeta(\Phi\cdot\Sigma)\bar{Y}^i
   +\zeta^2(\Sigma^j)^2\bar{Y}^i)\right\}
 \right]\,,
\end{eqnarray}
where
\begin{eqnarray}
 Y^i=L\Phi^i+\zeta\left\{L\Sigma^i-{\Phi^i \over r^2}
     \left(\bar{\Phi}\cdot\Sigma
     +{(\Phi\cdot\Sigma)(\bar{\Phi}^j)^2 \over 2r^2}\right)\right\}
     -\zeta^2 {\bar{\Phi}^i(\Sigma^j)^2 \over 2r^2}\,.
\end{eqnarray}
Here the denominator in the integrand 
 can be written as $\zeta^3g(\Phi,\Sigma)$ and the function of $\Phi$
 and $\Sigma$, $g(\Phi,\Sigma)$ can be factored out from the integral.
Recalling that $m\ge 2$, the numerator is a polynomial starting
 from an order of $\zeta^0$ in the $m=2$ case. 
In this case, the terms greater than the 
 first order of $\zeta$ vanish because of the
 residue theorem and the terms in the order of $\zeta^0$ only remain.
However, they cancel and the integral becomes zero.
In the case $m\ge 3$, the integral are trivially zero since all terms in
 the numerator starts from first order of $\zeta$.

Substituting the solution (\ref{sol-upsilon}), 
with the index $i$ running from $1$ to $n-2$, 
into the K{\"a}hler
 potential (\ref{kahler_SOn}), we can derive the nonlinear sigma model
 action with $TQ^{n-2}$ space:
\begin{eqnarray}
 S&=&\int d^8z
   {1 \over 2\pi i}
   \oint {d\zeta \over \zeta}\left[\Psi +\bar{\Psi}
    +r^2\ln\left(1+{|\Phi^i|^2 \over r^2}
     +{(\Phi^i)^2(\bar{\Phi}^{\bar{j}})^2 \over r^4} \right)\right. \nonumber \\
  &&
  +r^2\ln\left\{1-{1 \over r^2}g_{i\bar{j}}\Sigma^i\bar{\Sigma}^{\bar{j}}
     +{1 \over 2r^2}\left((g_{i\bar{j}}\Sigma^i\bar{\Sigma}^{\bar{j}})^2
     +{r^2 \over 2}R_{i\bar{j}k\bar{l}}\Sigma^i\bar{\Sigma}^{\bar{j}}
      \Sigma^k\bar{\Sigma}^{\bar{l}}\right)\right\}{\Bigg ]}\,,
\end{eqnarray}
with $\Psi$ defined by 
\begin{eqnarray}
 \Psi&=&-r^2\ln\left\{
        1-{\zeta \over r^2L}\left(\bar{\Phi}\cdot\Sigma+
        {(\Phi\cdot\Sigma)(\bar{\Phi}^i)^2 \over 2r^2}\right)
        +\zeta^2{(\bar{\Phi}^i)^2(\Sigma^j)^2 \over 4r^4L}\right\}\,.
\end{eqnarray}
After performing the $\zeta$-integration, we arrive at the same form
 with (\ref{action-tan}) with the index $i$ running from $1$ to $n-2$.

In Appendix \ref{so3}, 
we show this action in $n=3$  
coincides with the result in section 3 
due to isomorphism 
$Q^1\simeq {\bf C}P^1$. 

%%%%%%%%%%%%%%%%%%%%%%%%%%%%%%%%%%%%%%%%%%%%%%%%%%%%%%%%%%%%%%%5
\section{Cotangent bundle $T^* Q^{n-2}$}
Here we consider the nonlinear sigma model action with the 
 cotangent bundle 
 over $Q^{n-2}$, $T^*Q^{n-2}$ (hyper-K{\"a}hler potential).
To obtain the cotangent bundle, we need dualize the tangent vector
 $\Sigma^i$ into cotangent one $\psi_i$.
In order to do that, 
 let us consider the following Lagrangian instead of the tangent bundle
 action (\ref{action-tan}):
\begin{eqnarray}
  S&=&\int d^8z 
  \left[r^2\ln\left(1+{|\Phi^i|^2 \over r^2}
   +{(\Phi^i)^2(\bar{\Phi}^{\bar{j}})^2 \over 4r^4}\right)
   \right. \nonumber \\
 && 
  \left.
  +r^2\ln\left\{1-{1 \over r^2} g_{i\bar{j}}
    U^i\bar{U}^{\bar{j}}
  + {1 \over 2r^4}\left((g_{i\bar{j}}
    U^i\bar{U}^{\bar{j}})^2+{r^2 \over 2}
    R_{i\bar{j}k\bar{l}}U^i\bar{U}^{\bar{j}}
    U^k\bar{U}^{\bar{l}}\right)\right\}
  + U^i\psi_i+\bar{U}^i\bar{\psi}_{\bar{i}}\right]\,,\nonumber \\
 \label{legendre}
\end{eqnarray}
where $U^i$ and $\bar{U}^{\bar{j}}$ are 
unconstrained superfields, 
regarded as a tangent vector, 
and $\psi_i$ and $\bar{\psi}_{\bar{j}}$ are chiral superfields, 
regarded as a cotangent vector. 
Equation of motion of $\psi$ brings back 
to original Lagrangian (\ref{action-tan}). 
On the other hand, 
varying with respect to $U^i$ and $\bar{U}^{\bar{j}}$, we obtain  
\begin{eqnarray}
 \psi_i&=&{g_{i\bar{j}}\bar{U}^{\bar{j}}
        -{1 \over r^2}g_{i\bar{j}}\bar{U}^{\bar{j}}\rho
        -{1 \over 2}R_{i\bar{j}k\bar{l}}\bar{U}^{\bar{j}}
         U^k\bar{U}^{\bar{l}} \over 
        1-{\rho \over r^2}+{\rho^2 \over 2r^4}+{\sigma \over 4r^2}}\,, \label{aux1}\\
 \bar{\psi}_{\bar{j}}&=&{g_{i\bar{j}}U^i
        -{1 \over r^2}g_{i\bar{j}}U^i \rho
        -{1 \over 2}R_{i\bar{j}k\bar{l}}U^i
         U^k\bar{U}^{\bar{l}} \over 
        1-{\rho \over r^2}+{\rho^2 \over 2r^4}+{\sigma \over 4r^2}}\,, \label{aux2}
\end{eqnarray}
where  
\begin{eqnarray} 
 \rho\equiv g_{i\bar{j}}U^i\bar{U}^{\bar{j}} , \quad 
 \sigma\equiv R_{i\bar{j}k\bar{l}}U^i\bar{U}^{\bar{j}}
 U^k\bar{U}^{\bar{l}}.
\end{eqnarray}
Multiplying (\ref{aux1}) and (\ref{aux2}) by $U^i$ 
 and $\bar{U}^{\bar{i}}$, respectively, we get
\begin{eqnarray}
 U^i\psi_i=\bar{U}^{\bar{i}}\bar{\psi}_{\bar{i}}
 ={\rho-{\rho^2 \over r^2}-{\sigma \over 2} \over 1-{\rho \over r^2}
  +{\rho^2 \over 2r^4}+{\sigma \over 4r^2}}\,. \label{aux3}
\end{eqnarray}
Substituting Eqs.~(\ref{aux1}), (\ref{aux2}) and (\ref{aux3}) 
into the action (\ref{legendre}),
 one can see that the action (\ref{legendre}) is a function of
  the covariant quantities $\rho$ and $\sigma$ along with base manifold
  coordinates $\Phi^i$.
In the following, we show that $\rho$ and $\sigma$ are written by
 quantities
\begin{eqnarray}
 \xi\equiv  g^{i\bar{j}}\psi_i\bar{\psi}_{\bar{j}},~~~~
 \chi\equiv 
 R^{\bar{i}j\bar{k}l}\bar{\psi}_{\bar{i}}\psi_{j}\bar{\psi}_{\bar{k}}\psi_l,
 \label{invariant}
\end{eqnarray}  
 which are possible scalars in terms of cotangent vectors $\psi_i$ 
 and $\bar{\psi}_{\bar{i}}$.
In order to do that, first we substitute (\ref{aux1}) and (\ref{aux2})
 into Eq. (\ref{invariant}), and show that $\xi$ and $\chi$ are written
 in terms of $\rho$ and $\sigma$ with the help of the identities 
 between metric and Riemann tensor.
Then, solving them with respect to $\rho$ and $\sigma$, and substituting
 the solution into the action (\ref{legendre}), we obtain the 
 cotangent bundle action.
In the following, we give an explicit calculation on the 
 $T^*Q^2$ case.
The detailed derivation of the identities in this case is given
 in Appendix \ref{identities}. 
The resultant equations after substituting (\ref{aux1}) and (\ref{aux2})
 into (\ref{invariant}) is
\begin{eqnarray}
\xi=g^{i\bar{j}}\psi_i\bar{\psi}_{\bar{j}}
 &=&\left(1-{\rho \over r^2}+{\rho^2 \over 2r^4}+{\sigma \over 4r^2}\right)^{-2}
  \left(\rho-{2\rho^2 \over r^2}-\sigma+{\rho^3 \over 2r^4}
  +{\rho\sigma \over 4r^2}\right)\,, 
 \label{b4} \\
\chi=R^{\bar{i}j\bar{k}l}\bar{\psi}_{\bar{i}}\psi_{j}\bar{\psi}_{\bar{k}}\psi_l
 &=&\left(1-{\rho \over r^2}+{\rho^2 \over 2r^4}+{\sigma \over 4r^2}\right)^{-4} 
     \nonumber \\
 &&\times\left(\sigma+{2\rho\sigma \over r^2}-{6\rho^2\sigma \over r^4}
    +{2\rho^3\sigma \over r^6}
    +{\rho^4\sigma \over 4r^8}
    +{\sigma^3 \over 16r^4}+{\rho^2\sigma^2 \over 4r^6}-{6\rho^4 \over r^6}
    \right. \nonumber  \\
 &&\left.+{2\rho^5 \over r^8}-{3\sigma^2 \over 2r^2}+{\rho\sigma^2 \over 2r^4}
   +{4\rho^3 \over r^4}\right)\,. \label{c5}
\end{eqnarray}
Solving Eqs.~(\ref{b4}) and (\ref{c5}) 
 with respect to $\rho$ and $\sigma$ 
we obtain
\begin{eqnarray}
 \rho&=&2r^2\left(1
     -{1 \over 1+\sqrt{1+{2\xi \over r^2}
      +2\sqrt{-{\xi^2 \over r^4}-{\chi \over r^2}}}}
     -{1 \over 1+\sqrt{1+{2\xi \over r^2}
      -2\sqrt{-{\xi^2 \over r^4}-{\chi \over r^2}}}}
     \right)\,, \label{f1} \\
 \sigma&=&-4r^2\left\{1
     +{2 \over \left(1+\sqrt{1+{2\xi \over r^2}
      -2\sqrt{-{\xi^2 \over r^4}-{\chi \over r^2}}}\right)^2}
     +{2 \over \left(1+\sqrt{1+{2\xi \over r^2}
      +2\sqrt{-{\xi^2 \over r^4}-{\chi \over r^2}}}\right)^2} \right. 
      \nonumber \\
  & & \left.
      -{2 \over 1+\sqrt{1+{2\xi \over r^2}
      -2\sqrt{-{\xi^2 \over r^4}-{\chi \over r^2}}}}
\right\}\,. \label{f2}
\end{eqnarray}
Substituting back (\ref{f1}) and (\ref{f2}) into the action
 (\ref{legendre}), we obtain the action for the contangent bundle 
 $T^*Q^{2}$ 
\begin{eqnarray}
 S&=&\int d^8 z {\Bigg \{}r^2\ln\left(1+{|\Phi^i|^2 \over r^2}
    +{(\Phi^i)^2(\bar{\Phi}^{\bar{j}})^2 \over 4r^2}\right) 
    \nonumber \\
   && -r^2\ln\left({1\over 2}+{1\over 2}\sqrt{1+{2\xi \over r^2}
      +2\sqrt{-{\xi^2 \over r^4}-{\chi \over r^2}}}\right) 
      -r^2\ln\left({1\over 2}+{1\over 2}\sqrt{1+{2\xi \over r^2}
      -2\sqrt{-{\xi^2 \over r^4}-{\chi \over r^2}}}\right)
     \nonumber \\
   &&\left.+{\xi
      +r^2\sqrt{-{\xi^2 \over r^4}-{\chi \over r^2}} \over
       {1\over 2}+{1\over 2}\sqrt{1+{2\xi \over r^2}
      +2\sqrt{-{\xi^2 \over r^4}-{\chi \over r^2}}}}
     +{\xi
      -r^2\sqrt{-{\xi^2 \over r^4}-{\chi \over r^2}} \over
      {1\over 2}+{1\over 2}\sqrt{1+{2\xi \over r^2}
      -2\sqrt{-{\xi^2 \over r^4}-{\chi \over r^2}}}}
\right\}\,.\label{action-cot} 
\end{eqnarray}
This action is of $T^*Q^2$.
When one tries to derive the cotangent bundle action 
of $T^*Q^{n-2}$ case, 
one comes across difficulty 
in deriving the identities
for cotangent bundle action for $T^*Q^{n-2}$,  
as shown in Appendix \ref{identities}.
However, we believe that this form of the action 
 would be also valid in $TQ^{n-2}$ case with 
extending the index
 to run from $1$ to $n-2$, 
 since the action is written in geometric
 quantities such as K{\"a}hler potential, metric and Riemann tensor 
as in the tangent bundle case.

One might consider  
 a possibility that  
 invariant quantities higher in 
$\psi$ 
such like
 $(g^{i\bar{j}}\psi_i\bar{\psi}_{\bar{j}})^3$
 and 
 $(g^{i\bar{j}}\psi_i\bar{\psi}_{\bar{j}})
  (R^{\bar{k}l\bar{m}n}\bar{\psi}_{\bar{k}}\psi_{l}\bar{\psi}_{\bar{m}}\psi_n)$
 together with $\xi$
 and $\chi$ appear in the cotangent bundle action.
However,
 one can confirm 
 by deriving the cotangent bundle action in another way
 that such terms do not appear. 
We derive the action (\ref{action-cot}) with the similar method
 used in subsection \ref{TCP1}.
Using the isomorphism $T^*Q^2\simeq T^*{\bf C}P^1\times T^*{\bf C}P^1$,
 we can represent the $T^*Q^2$ action as a sum of
 two $T^*{\bf C}P^1$ action with different coordinate variables.
Although naive summation is not written by geometrical quantities, 
 similarly to the tangent bundle case, we can write the action 
 in terms of metric and Riemann tensor.
Finally we can reach the same form with (\ref{action-cot}) which
 is written in terms on $\xi$ and $\chi$ with base manifold coordinates.
It is shown in Appendix \ref{cotangent}.
Discussion on $n=3$ case is given in Appendix \ref{so3}.

%%%%%%%%%%%%%%%%%%%%%%%%%%%%%%%%%%%%%%%%%%%%%%%
\section{Discussion}
We would like to propose possible future works. 
Potential terms in ${\cal N}=2$ hyper-K\"ahler nonlinear sigma models 
are known to be written in the form of the square of 
a Killing vector of the target space 
\cite{Alvarez-Gaume:1983ab,Bagger:2006hm}. 
Recently these potentials in the framework of 
the projective superspace has been 
established \cite{Kuzenko:2006nw} by 
gauging the isometry to obtain the potential 
and then by freezing it. 
Using this method we could construct the potential 
of our $T^{(*)} Q^n$ nonlinear sigma model. 
Massive extension of 
$T^* {\bf C}P^n $\cite{Arai:2002xa} 
and $T^*$ Grassmann \cite{Arai:2003tc}
nonlinear sigma models were formulated 
by the hyper-K\"ahler quotient
in the harmonic superspace formalism as well as 
components and ${\cal N}=1$ superfields. 
These models are known to admit various (composite) BPS solitons, like 
domain walls \cite{walls,Arai:2002xa,Isozumi:2004jc,Eto:2005wf}, 
(Q-)lump-strings \cite{lump} (vortex-strings \cite{vortices}), 
domain wall junctions \cite{webs}, 
strings stretched between walls \cite{Isozumi:2004vg,wvm}, 
intersecting vortex-strings \cite{ivv,Naganuma:2001pu}, 
and other solitons \cite{other}. 
See \cite{Tong:2005un,review2,review3} for a review 
in this subject.
So constructing 
the massive  $T^* Q^n$ model 
and investigating BPS solitons in it 
are interesting future directions.

We have seen in the Lagrangian (\ref{Q^n-lag}) in the introduction 
that the ${\cal N}=1$ $Q^n$ model 
can be constructed by the K\"ahler quotient.  
It has been found \cite{Isozumi:2004jc,Eto:2005wf,vortices,webs,ivv}
that the hyper-K\"ahler quotient construction 
is crucial in solving the BPS equations to construct 
all the {\it exact} soliton solutions.  
Therefore, hyper-K\"ahler quotient construction 
of the (massive)  $T^* Q^n$ model is awaited. 
${\cal N}=1$ superfield formalism is 
difficult to perform it because 
the superpotential exists 
even in the massless case (\ref{Q^n-lag}). 
The projective superspace formalism 
should be useful to construct 
the massive  $T^* Q^n$ model in hyper-K\"ahler quotient.

The cotangent bundle over the projective space 
${\bf C}P^{n-1}$ can be locally written as 
\begin{eqnarray}
 T^* {\bf C}P^{n-1} 
 =  T^* \left[SU(n) \over SU(n-1)\times U(1) \right]
 \simeq {\bf R} \times {SU(n)\over SU(n-2)}. 
\end{eqnarray}
Therefore, this is in cohomogeneity one. 
It was proved \cite{cohomo1} that 
this is the unique cohomogeneity one hyper-K\"ahler manifold.
On the other hand, $T^* Q^{n-2}$ can be locally written as
\begin{eqnarray}
 T^* Q^{n-2} 
 =  T^* \left[SO(n) \over SO(n-2)\times U(1)\right]
 \simeq {\bf R}^2 \times {SO(n)\over SO(n-4)} 
\end{eqnarray}
which is in cohomogeneity two. 
Cohomogeneity two hyper-K\"ahler manifolds were discussed
in \cite{cohomo2} but complete classification is not yet known 
at least to our knowledge. 
We hope our example is useful to explore 
hyper-K\"ahler manifolds with higher cohomogeneity.

There exist other 
Hermitian symmetric spaces,  
$G_{n,m} = SU(n)/[SU(n-m) \times SU(m) \times U(1)]$, 
$SO(2n)/U(n)$, $Sp(n)/U(n)$, $E_6/SO(10) \times U(1)$ 
 and $E_7/E_6 \times U(1)$.  
A K\"ahler quotient construction of these manifolds was
given in
\cite{Higashijima:1999ki}.\footnote{
The Calabi-Yau metrics on line bundles over Hermitian symmetric 
spaces were constructed \cite{Higashijima:2001de,Higashijima:2001vk}, 
which are all in cohomogeneity one.
}
It should be possible to construct 
(co)tangent bundles over Hermitian symmetric spaces \cite{BG}
in the framework of the projective superspace. 
Extension to (co)tangent bundle over 
arbitrary homogeneous K\"ahler manifold $G/H$  
is one goal of this subject.  
To this end, tangent bundles 
are in more compact form than cotangent bundles 
as seen in the cases of ${\bf C}P^n$ (\ref{final_CP}) 
and $Q^n$ (\ref{action-tan}). 
General form of the K\"ahler potential 
of tangent bundle over arbitrary K\"ahler manifold \cite{Kaledin,Feix}
expanded in terms of a tangent vector $\Sigma$
was proposed in \cite{kuzenko,kuzenko2}. 
There, each coefficient should be written in terms of 
the metric, curvature, covariant derivative like 
the cases of ${\bf C}P^n$ (\ref{final_CP}) 
and $Q^n$ (\ref{action-tan}), 
but explicit expression is not known in general. 
This expansion is very similar with  
the K\"ahler normal coordinate expansion \cite{Higashijima:2000wz}. 
We hope that K\"ahler normal coordinates 
are useful toward the construction of 
general action of hyper-K\"ahler 
nonlinear sigma models on (co)tangent bundle 
over arbitrary K\"ahler manifold.
\vspace{5mm}\\\\
%%%%% Acknowledgements  %%%%%%
\noindent {\Large \bf Acknowledgements}

We would like to thank Rikard von Unge
 for useful discussions and comments, especially,  
 on duality between polar and $O(2)$ multiplets.
We thank Kiyoshi Higashijima and Ulf Lindstr\"om for discussions.
M.~A. acknowledges the Institute of Physics in Prague and the 
 Masaryk university for their hospitality 
while M.~N. is grateful to 
the Helsinki Institute of Physics for their hospitality.
The work of M.~A. is supported by the bilateral program of Japan Society 
 for the Promotion of Science and Academy of Finland, ``Scientist Exchanges'' 
while the work of M.~N. is 
supported by Japan Society for the Promotion 
of Science under the Post-doctoral Research Program.

%%%%%%%%%%%

\vspace{10mm}
\renewcommand{\thesubsection}{\thesection.\arabic{subsection}}
\noindent{\Large \bf Appendix}
\appendix
\section{Duality between polar and $O(2)$ multiplets}
Here we show the other way to eliminate infinite number of 
 auxiliary fields, which is not mentioned in the text.
The point is to perform a duality transformation 
 \cite{rocek,deWit:2001dj} between the polar
 multiplet $\Upsilon$ and real $O(2)$ multiplet defined by
\begin{eqnarray}
 \eta=\bar{\Phi}+\zeta\Sigma-\zeta^2\Phi\,,~~~~\Sigma=\bar{\Sigma}\,.
\end{eqnarray}
The duality transformation is possible only if the polar multiplet
 appears as the linear combination such that $\Upsilon+\bar{\Upsilon}$ in
 an action.
After the duality transformation, the action is described by $O(2)$
 multiplet and there is thus no auxiliary fields.
\footnote{As more general case, there is duality between the real $O(2p)$ and
 polar multiplets \cite{gon0}, 
 which involves the case we will discuss below.}  
Let us illustrate this with a couple of examples.

Firstly we consider the flat space. 
Let us start
with an action depending on the arctic projective 
 multiplet $\Upsilon$:
\be
S = \frac{1}{2}\int d^8z \oint \frac{d\zeta}{2\pi i \zeta} 
 \left(\Upsilon+\bar{\Upsilon}\right)^2\,.
\ee
This is a sigma model in flat space since the Lagrangian consists of 
 the K\"ahler potential $\Upsilon\bar{\Upsilon}$ of flat space
 and a K\"ahler transformation.
Now we instead introduce an $O(2)$ multiplet $\eta$ and an 
 unconstrained projective superfield $X$
\be
S = \int d^8z \oint \frac{d\zeta}{2\pi i \zeta}
\left( \frac12 X^2-X\frac{\eta}{\zeta}\right)\,.
\ee
Varying with respect to $\eta$ tells us that $X$ can be written as a sum
 of an arctic and antarctic field $X = \Upsilon + \bar{\Upsilon}$ and we
 are back to the original model. If we on the other hand vary with
 respect to $X$ we get an equation
\be
X = \frac{\eta}{\zeta}\,.
\ee
Inserting this back into the action we get
\be
S=-\int d^8z \oint \frac{d\zeta}{2\pi i \zeta} \frac{\eta^2}{\zeta^2}\,,
\ee
which is the action for a sigma model in flat space written with an ${O}(2)$ multiplet.

In the second example, 
 we start from another K\"ahler potential 
 in flat space:
\be
S = \int d^8z\oint \frac{d\zeta}{2\pi i \zeta} e^{\Upsilon+\bar{\Upsilon}}.
\ee
In the same way, we introduce an $O(2)$ multiplet $\eta$ and 
 a polar multiplet $X$ to get the action
\be
S = \int d^8z\oint \frac{d\zeta}{2\pi i \zeta} 
 \left(e^{X} - X\frac{\eta}{\zeta}\right)\,.
\ee
Again, integrating out $\eta$ tells us that $X$ can be written 
 as a sum over an arctic and an antarctic multiplet
  while integrating out $X$ gives %you 
the action
\be
S = \int d^8z\oint \frac{d\zeta}{2\pi i \zeta} 
 \frac{\eta}{\zeta}\left(1-\ln\frac{\eta}{\zeta}\right)\,.
\ee
This is also a known form of the action for a sigma model in flat
 space. There is one unresolved issue here with the integration
 contour. Because of the nonsingle valuedeness of the log, the
 integration contour will not be simply a contour around the origin but
 rather a shaped curve around the zeroes of the 
 quadratic polynomial $\eta(\zeta)$.

This method can be applied to any K\"ahler potential which can be 
 written as a function of $\Upsilon+\bar{\Upsilon}$. Namely
\be
S = \int d^8z\oint \frac{d\zeta}{2\pi i \zeta} 
 f(\Upsilon+\bar{\Upsilon}) =
 \int d^8z\oint \frac{d\zeta}{2\pi i \zeta} 
 \left(f(X)-X\frac{\eta}{\zeta}\right)\,.
\ee
Integrating out $X$ one gets
\be
f'(X)=\frac{\eta}{\zeta}\,,
\ee
which can be formally inverted to give
\be
X = g\left(\frac{\eta}{\zeta}\right)\,.
\ee
Inserting this back into the action gives
\be
S = \int d^8z\oint \frac{d\zeta}{2\pi i \zeta} 
 \left\{f\left(g\left(\frac{\eta}{\zeta}\right)\right)
 -\frac{\eta}{\zeta}g\left(\frac{\eta}{\zeta}\right)\right\}\,.
\ee
Technically we see that the new action is the Legendre transform of the
 old one. 
For example, starting with the K\"ahler potential 
 of $T^*{\bf C}P^1$ $\ln\left(1+e^{\Upsilon+\bar{\Upsilon}}\right)$, we get
\be
S &=& \int d^8z\oint \frac{d\zeta}{2\pi i \zeta} 
 \ln\left(1+e^{\Upsilon+\bar{\Upsilon}}\right)
\nonumber\\ &=&\int d^8z\oint \frac{d\zeta}{2\pi i \zeta}
\left(-\frac{\eta}{\zeta}\ln\left(\frac{\eta}{\zeta}\right)-
\left(1-\frac{\eta}{\zeta}\right)\ln\left(1-\frac{\eta}{\zeta}\right)\right)\,.
\ee
which is again a known form for the nonlinear sigma model with 
 the Eguchi-Hanson metric. 
The contour is still an issue similarly to the flat 
 case $e^{\Upsilon+\bar{\Upsilon}}$.
As be seen,
 one sees how one does not have to 
 solve the infinite number of equations
 although it is not applicable to the quadric surface.
 
\section{Deriving equations (\ref{b4}) and (\ref{c5})}\label{identities}
In this Appendix, we show how to derive the equations
 (\ref{b4}) and (\ref{c5}).
Substituting (\ref{aux1}) and (\ref{aux2}) into $\xi$ and $\chi$ 
 leads to, after some algebra
\begin{eqnarray}
\xi&=&g^{i{\bar{j}}}\psi_i\bar{\psi}_j \nonumber \\
 &=&\left(1-{\rho \over r^2}+{\rho^2 \over 2r^4}+{\sigma \over 4r^2}\right)^{-2} 
     \nonumber \\
 &&\times\left(\rho-{2\rho^2 \over r^2}-\sigma+{\rho^3 \over r^4}
    +{\rho\sigma \over r^2}
    +{g^{i\bar{j}} \over 4}R_{i\bar{l}p\bar{q}}R_{k\bar{j}m\bar{n}}
    \bar{U}^{\bar{l}}U^p\bar{U}^{\bar{q}}
    U^kU^m\bar{U}^{\bar{n}}
 \right)\,. \label{b0} \\
\chi&=&R^{\bar{i}j\bar{l}k}\bar{\psi}_{\bar{i}}\psi_{j}\bar{\psi}_{\bar{l}}\psi_k
 \nonumber \\
 &=&\left(1-{\rho \over r^2}+{\rho^2 \over 2r^4}+{\sigma \over 4r^2}\right)^{-4}
     R^{\bar{j}i\bar{l}k} \nonumber \\
 &&\times{\Bigg [}
  g_{n\bar{j}}g_{i\bar{m}}g_{a\bar{l}}g_{k\bar{b}}
  U^n\bar{U}^{\bar{m}}U^a\bar{U}^{\bar{b}}
  \left(1-{4\rho \over r^2}+{6\rho^2 \over r^4}
  -{4\rho^3 \over r^6}+{\rho^4 \over r^8}\right) \nonumber \\
 &&+ g_{n\bar{j}}g_{i\bar{m}}g_{k\bar{a}}R_{b\bar{l}c\bar{d}}
  U^n\bar{U}^{\bar{m}}\bar{U}^{\bar{a}}U^bU^c
  \bar{U}^{\bar{d}}\left(-2+{6\rho \over r^2}-{6\rho^2 \over r^4}
       +{2\rho^3 \over r^6} \right) \nonumber \\
 &&+ g_{i\bar{m}}g_{k\bar{a}}R_{n\bar{j}p\bar{q}}R_{b\bar{l}c\bar{d}}
   \bar{U}^{\bar{m}}U^nU^p\bar{U}^{\bar{q}}
   \bar{U}^{\bar{a}}U^bU^c\bar{U}^{\bar{d}}
   \left({3 \over 2}-{3\rho \over r^2}+{3\rho^2 \over 2r^4}\right) \nonumber \\
 &&+{1 \over 2}
   g_{k\bar{x}}R_{i\bar{m}n\bar{p}}R_{a\bar{j}b\bar{c}}R_{y\bar{l}z\bar{w}}
   \bar{U}^{\bar{x}}\bar{U}^{\bar{m}}U^n\bar{U}^{\bar{p}}
   U^aU^b\bar{U}^{\bar{c}}U^yU^z\bar{U}^{\bar{w}}
   \left(-1+{\rho \over r^2}\right) \nonumber \\
 &&+{1 \over 16}R_{i\bar{m}n\bar{p}}R_{a\bar{j}b\bar{c}}R_{k\bar{x}y\bar{z}}
   R_{d\bar{l}e\bar{f}}\bar{U}^{\bar{m}}U^n\bar{U}^{\bar{p}}
   U^aU^b\bar{U}^{\bar{c}}\bar{U}^{\bar{x}}U^y
   \bar{U}^{\bar{z}}U^dU^e\bar{U}^{\bar{f}}
 {\Bigg ]}\,. \label{b6}
\end{eqnarray}
It is seen that these equations are not closed with respect to $\rho$
 and $\sigma$.
However, using identities between metric and Riemann tensor, one can
 rewrite terms such like 
 $g^{i\bar{j}}R_{i\bar{l}p\bar{q}}R_{k\bar{j}m\bar{n}}
  \bar{U}^{\bar{l}}U^p\bar{U}^{\bar{q}}U^kU^m\bar{U}^{\bar{n}}$
 in the right hand side in (\ref{b0}) in terms of
 $\rho$ and $\sigma$.

First we consider Eq. (\ref{b0}).
Calculating last term in the right hand side in (\ref{b0})
 in the frame taken in (\ref{action-base}), we find
\begin{eqnarray}
 g^{i\bar{j}}R_{i\bar{l}p\bar{q}}R_{k\bar{j}m\bar{n}}
 \bar{U}^{\bar{l}}U^p\bar{U}^{\bar{q}}U^kU^m
 \bar{U}^{\bar{n}}
 ={4|U^1|^6 \over r^4 \left(1+{|\Phi_1|^2 \over r^2}\right)^6}
 +{4|U^2|^6 \over r^4 \left(1+{|\Phi_2|^2 \over r^2}\right)^6}\,.
 \label{b1}
\end{eqnarray}
On the other hand, it is shown that
\begin{eqnarray}
 \rho\sigma&=&-{2|U^1|^6 \over r^2 \left(1+{|\Phi_1|^2 \over r^2}\right)^6}
    -{2|U^2|^6 \over r^2 \left(1+{|\Phi_2|^2 \over r^2}\right)^6} \nonumber \\
   &&-{2 \over r^2}G_1G_2|U^1|^2|U^2|^2
     (G_1|U^1|^2+G_2|U^2|^2)\,. \label{b2}
\end{eqnarray}
Using (\ref{b1}), (\ref{b2}) and (\ref{a3}) with 
 $\rho=g_{i\bar{j}}U^i\bar{U}^{\bar{j}}$, we get the following identity
\begin{eqnarray}
 g^{i\bar{j}}R_{i\bar{l}p\bar{q}}R_{k\bar{j}m\bar{n}}
 \bar{U}^{\bar{l}}U^p\bar{U}^{\bar{q}}U^kU^m
 \bar{U}^{\bar{n}}
 =-{3\rho\sigma \over r^2}-{2 \over r^4}\rho^3\,. \label{b3}
\end{eqnarray}
Since this identify is written by geometric quantities, it holds in
 arbitrary frame.
Substituting (\ref{b3}) into (\ref{b0}), we obtain (\ref{b4}).

Next we consider Eq. (\ref{b6}).
Once again choosing the frame taken in (\ref{action-base}), we find
 the following identities:
\begin{eqnarray}
 R^{\bar{i}j\bar{k}l}g_{n\bar{j}}g_{i\bar{m}}
 g_{i\bar{m}}g_{k\bar{a}}R_{n\bar{j}p\bar{q}}R_{b\bar{l}c\bar{d}}
   \bar{U}^{\bar{m}}U^nU^p\bar{U}^{\bar{q}}
   \bar{U}^{\bar{a}}U^bU^c\bar{U}^{\bar{d}}&=&
 {4\rho^4 \over r^6}-{\sigma^2 \over r^2}+{4\rho^2 \sigma \over r^4}\,, 
 \label{c2} \\
 R^{\bar{i}j\bar{k}l}g_{n\bar{j}}g_{i\bar{m}}
 g_{k\bar{x}}R_{i\bar{m}n\bar{p}}R_{a\bar{j}b\bar{c}}R_{y\bar{l}z\bar{w}}
   \bar{U}^{\bar{x}}\bar{U}^{\bar{m}}U^n\bar{U}^{\bar{p}}
   U^aU^b\bar{U}^{\bar{c}}U^yU^z\bar{U}^{\bar{w}}
 &=&{5\rho\sigma^2 \over r^4}-{4\rho^5 \over r^8}\,,\label{c3} \\
 R^{\bar{i}j\bar{k}l}g_{n\bar{j}}g_{i\bar{m}}
 R_{i\bar{m}n\bar{p}}R_{a\bar{j}b\bar{c}}R_{k\bar{x}y\bar{z}}
   R_{d\bar{l}e\bar{f}}\bar{U}^{\bar{m}}U^n\bar{U}^{\bar{p}}
   U^aU^b\bar{U}^{\bar{c}}\bar{U}^{\bar{x}}U^y
   \bar{U}^{\bar{z}}U^dU^e\bar{U}^{\bar{f}}
 &=&{\sigma^3 \over r^4}-{12 \rho^4\sigma \over r^8}
 -{12 \rho^2\sigma^2 \over r^6}\,. \nonumber \\ \label{c4}
\end{eqnarray}
Substituting (\ref{b3}) and (\ref{c2})-(\ref{c4}) into (\ref{b6}), we find the 
 equation (\ref{c5}).

\section{Derivation of cotangent bundle without using Legendre
  transformation}\label{cotangent}
Instead of using the Legendre transformation, we can directly obtain
 the cotangent bundle action (\ref{action-cot}) for $T^*Q^2$ case.
Similarly to the tangent bundle case,
 with the isomorphism $T^*Q^2=T^*{\bf C}P^1\times T^*{\bf C}P^1$,
 the action for $T^*Q^2$ can be written as a direct sum of
 two actions of cotangent bundle over ${\bf C}P^1$.
Using the action (\ref{cot-cp}), we have
\begin{eqnarray}
 S&=&\int d^8 z {\Bigg \{}r^2\ln\left(1+{|\Phi^i|^2 \over r^2}
 +{|\Phi^1|^2|\Phi^2|^2 \over r^4}\right)
    \nonumber \\
   &&\left. -r^2\ln f(\kappa_1)+2r^2{\kappa_1 \over f(\kappa_1)}
   -r^2\ln f(\kappa_2)+2r^2{\kappa_2 \over f(\kappa_2)}
\right\}\,, \label{cot-Q2}
\end{eqnarray}
where 
\begin{eqnarray}
 f(\kappa_i)={1 \over 2}(1+\sqrt{1+4\kappa_i})\,,~~~~~
 \kappa_i={1 \over r^2}G_i^{-1}\psi_i\bar{\psi}_{\bar{i}}\,,~~~(\mbox{no sum}).
\end{eqnarray}
Here $G_i$ is defined in (\ref{metriccp}).
In the following, 
 we rewrite the action (\ref{cot-Q2}) in terms of 
 geometric quantities
 similarly to the tangent bundle case.
We consider the two quantities
 $\xi=g^{i\bar{j}}\psi_i\bar{\psi}_{\bar{j}}$ and 
 $\chi=R^{\bar{i}j\bar{k}l}\bar{\psi}_{\bar{i}}\psi_j\bar{\psi}_{\bar{k}}\psi_l$
 as possible invariant forms written by 
 $\psi_i$ and $\bar{\psi}_{\bar{j}}$
 as the same with in section 5.
In the frame taken in (\ref{action-base}), one finds
 \begin{eqnarray}
 \xi&=&{1 \over G_1}|\psi_1|^2+
  {1 \over G_2}|\psi_2|^2\,, \label{d1} \\
 \chi
 &=&-{2 \over G_1^2r^2}|\psi_1|^4-{2 \over G_2^2r^2}|\psi_2|^4\,. \label{d2}
\end{eqnarray}
Using (\ref{d1}) and (\ref{d2}) leads to
\begin{eqnarray}
&\displaystyle\xi^2={|\psi_1|^4 \over G_1^2}+{|\psi_2|^4 \over G_2^2}
    +{2 \over G_1G_2}|\psi_1|^2|\psi_2|^2=-{r^2 \over 2}
 R^{\bar{i}j\bar{k}l}\bar{\psi}_{\bar{i}}\psi_j\bar{\psi}_{\bar{k}}\psi_l
 +{2 \over G_1G_2}|\psi_1|^2|\psi_2|^2\,, \nonumber &\\
&\displaystyle\Leftrightarrow {1 \over r^2 G_1G_2}|\psi_1|^2|\psi_2|^2
 ={\xi^2 \over 2r^4}
  +{\chi \over 4r^2}\,. \label{d3}&
\end{eqnarray}
From (\ref{d1}), (\ref{d2}) and (\ref{d3}), we get two equations
\begin{eqnarray}
 \kappa_1+\kappa_2&=&{\xi \over r^2}\,, \label{e1} \\
 \kappa_1\kappa_2&=&{\xi^2 \over 2r^4}+{\chi \over 4r^2}\,. \label{e2}
\end{eqnarray} 
Solving (\ref{e1}) and (\ref{e2}) with respect to $\kappa_1$ and
  $\kappa_2$, we obtain
\begin{eqnarray}
 (\kappa_1,\kappa_2)=
 \left({\xi \over 2r^2}
 +{1\over 2}\sqrt{-{\xi^2 \over r^4}-{\chi \over r^2}}\,, 
 {\xi \over 2r^2}
 -{1\over 2}\sqrt{-{\xi^2 \over r^4}-{\chi \over r^2}}\right)\,. \label{e3}
\end{eqnarray}
Substituting back this solution into the action (\ref{cot-Q2}) and
 performing the unitary transformation (\ref{unitary}), 
 we arrive at the action (\ref{action-cot}).
 
\section{$T^{(*)}Q^1$ case}\label{so3}
Here 
 we consider the $n=3$ case in the sigma model with $T^{(*)}Q^{n-2}$, 
 say, $T^{(*)}[SO(3)/SO(2)]\simeq T^{(*)}{\bf C}P^1$.
In this case, everything is drastically simplified.
The K{\"a}hler potential (\ref{kahler_SO})
 in this case reduces to the ${\bf C}P^1$'s one: 
\begin{eqnarray}
 K(\Phi^i,\bar{\Phi}_j)=\ln\left(1+{|\Phi|^2 \over r^2}
 +{(\Phi)^2(\bar{\Phi})^2 \over 4r^4} \right)
 =2\ln\left(1+{|\Phi|^2 \over 2r^2}\right)\,.
\end{eqnarray}
Eqs. (\ref{def1})--(\ref{def3}) become
\begin{eqnarray}
\lambda&=&\epsilon\bar{\epsilon}\,, \\
\beta_+&=&0,~~\beta_-=\sqrt{2\epsilon\bar{\epsilon}}\,, \\
\gamma_{+}&=&2\epsilon\lambda,~~\gamma_-=0\,. 
\end{eqnarray}
Now the 
 transformation factor $V_Q$ in Eq. (\ref{sigma-so}) can be 
 simply written as 
\begin{eqnarray}
 \Sigma=V_Q\Sigma_0=\left(1+{|\Phi|^2 \over 2r^2}\right)\Sigma_0\,.
\end{eqnarray}
Substituting these expressions into (\ref{upsilon3-so}), we find 
\begin{eqnarray}
 \Upsilon={\left(1+{|\Phi|^2 \over 2r^2}\right)\Phi + \zeta\Sigma \over
           1+{|\Phi|^2 \over 2r^2}-{\zeta \over 2r^2}\bar{\Phi}\Sigma}\,.
\end{eqnarray}
This result coincides with one of (\ref{upsilon-CP}) in the $n=2$ case
 with rescaling $r\rightarrow r/\sqrt{2}$.

We can also see that the actions (\ref{action-tan})
 and (\ref{action-cot}) reduce to ones of $T{\bf C}P^1$ 
 and $T^*{\bf C}P^1$ in the $n=3$ case.
In this case, the covariant quantities $\rho,\sigma,\xi$ and $\chi$ become
\begin{eqnarray}
 &\rho\rightarrow G|\Sigma|^2\,,~~~~
  \sigma\rightarrow -{1 \over r^2}G^2|\Sigma|^4\,,& \\
 &\xi\rightarrow G^{-1}|\psi|^2\,,~~~~
 \chi\rightarrow -{1 \over r^2}G^{-2}|\psi|^4\,,&
\end{eqnarray}
and Eqs. (\ref{e3}) and $f(\kappa_i)$ are
\begin{eqnarray}
 &\kappa_1\,,\kappa_2 \rightarrow {|\psi|^2 \over 2r^2 G}\,,&\\
 &f(\kappa_1)\,,f(\kappa_2)\rightarrow f(\kappa)\,,&
\end{eqnarray}
where $G=\left(1+|\Phi|^2/(2r^2)\right)^{-1}$ and 
 $f(\kappa)={1 \over 2}\left(1+\sqrt{1+{2|\psi|^2 \over r^2G}}\right)$.
Using these formulas, we find that the actions for tangent
 and cotangent bundles coincide with (\ref{final_CP}) and (\ref{cot-cp})
 in the $n=2$ case, respectively, with
 rescaling $r \rightarrow r/\sqrt{2}$.

%%%%%%%%%%%%%%%%%%%%%%%%%%%%%%%%%%%%%%%%%%%%%%%%%

\end{document}